\documentclass[11pt]{article}

\usepackage{amssymb,amsmath,amsfonts,amsthm}

 \usepackage[numbers,sort&compress]{natbib}
\usepackage{graphicx}

\usepackage{hyperref}
\hypersetup{
   colorlinks   =  true,
    linkcolor    = blue,
    citecolor    = blue,
     urlcolor	=blue,     
 }

\newcommand\be{\begin{equation}}
\newcommand\ee{\end{equation}}
\newcommand\bea{\begin{eqnarray}}
\newcommand\eea{\end{eqnarray}}

\theoremstyle{plain}

\theoremstyle{definition}

\numberwithin{theorem}{section}
\numberwithin{equation}{section}

\def\vv{\tilde  v}

\def\>#1{{\bf #1}}

\def\dd{{\rm d}}
 \def\lh{{\cal H}_\omega}
  
  \def\qlhz{   C^\infty({\cal H}_{z,\omega}^\ast) }
  
   \def\qlhzd{   C^\infty({\cal H}^*_{z,\omega}) }

\def\1{\'{\i}}                           
\def\mm{M} 
 
 \def\mmm{{\cal M}} 
 
 \def\h4{ \mathfrak{h}_{4,\omega} } 
  \def\hz{ \mathfrak{h}_{z4,\omega} }

\def\kk{k} 

\def\eee{{\rm e}}

\def\sss{l}

\begin{document}

\
  \vskip0.5cm

 \noindent
 {\Large \bf  
Poisson--Hopf deformations of Lie--Hamilton systems revisited:\\[5pt]  deformed superposition rules and  applications to the oscillator\\[5pt]  algebra}

\medskip 
\medskip 
\medskip

\begin{center}

{\sc Angel Ballesteros$^1$, Rutwig Campoamor-Stursberg$^{2,3}$, Eduardo Fern\'andez-Saiz$^{3}$,\\[4pt]   Francisco J.~Herranz$^1$ and Javier de Lucas$^{4}$}

\end{center}

\medskip

\noindent
{$^1$ Departamento de F\'isica, Universidad de Burgos, 
E-09001 Burgos, Spain}

 \noindent
$^2$ Instituto de Matem\'atica Interdisciplinar UCM, Plaza de Ciencias 3, E-28040
Madrid, Spain

\noindent 
$^3$ Departamento de Geometr\'{\i}a y Topolog\'{\i}a,  Universidad Complutense de Madrid, Plaza de Ciencias 3, E-28040 Madrid, Spain

\noindent
$^4$  Department of Mathematical Methods in Physics, University of Warsaw, ul. Pasteura 5, 02-093, Warszawa, Poland

 \medskip
 
\noindent  E-mail: {\small {\href{mailto:angelb@ubu.es}{angelb@ubu.es}, \href{mailto:rutwig@ucm.es}{rutwig@ucm.es}, \href{mailto:eduardfe@ucm.es}{eduardfe@ucm.es}, \href{mailto:fjherranz@ubu.es}{fjherranz@ubu.es}, \href{mailto:javier.de.lucas@fuw.edu.pl}{javier.de.lucas@fuw.edu.pl}
}

\medskip

\begin{abstract}
\noindent The formalism for Poisson--Hopf (PH) deformations of Lie--Hamilton systems, recently proposed in~\cite{BCFHL}, is refined in one of its crucial points concerning applications, namely the obtention of effective and computationally feasible PH deformed superposition rules for prolonged PH deformations of Lie--Hamilton systems. The two new notions here proposed are a generalization of  the   standard superposition rules and the concept of diagonal prolongations for Lie systems,  which are consistently recovered under the non-deformed limit.
  Using a technique from superintegrability theory, we obtain a maximal number of functionally independent constants of the motion for a generic prolonged PH deformation of a Lie--Hamilton system, from which a simplified  deformed superposition rule can be derived. As an application, explicit deformed superposition rules for prolonged PH deformations of Lie--Hamilton systems based on the oscillator Lie algebra $\mathfrak{h}_4$ are computed. Moreover, by making use that the main structural properties of the book subalgebra $\mathfrak{b}_2$ of $\mathfrak{h}_4$  are preserved under the PH deformation, we consider prolonged PH deformations based on  $\mathfrak{b}_2$  as restrictions of those for    $\mathfrak{h}_4$-Lie--Hamilton systems, thus allowing the study of prolonged PH deformations of the complex Bernoulli equations, for which both the constants of the motion and the deformed superposition rules are explicitly presented. 
 
  \end{abstract}
\medskip
\medskip


\medskip


\medskip

 \newpage

\tableofcontents


\section{Introduction}

  The original approach to Poisson--Hopf (PH) deformations of Lie--Hamilton systems  developed in~\cite{BCFHL} combined the classical theory of Lie systems with methods from quantum algebras and integrable systems, leading to a novel type of systems of ordinary differential equations with generalized symmetry that, despite being deprived of some of the appealing properties of Lie systems, still allowed for a systematic analysis of their constants of the motion. In essence, the method is based on the idea of deforming a Lie--Hamilton system (LH system in short) with a given Vessiot--Guldberg Lie algebra  onto a Hamiltonian system depending on a quantum deformation parameter $z$ (or $q={\rm e}^z$), the dynamics of which is described by a $t$-dependent vector field taking values in a linear space of vector fields spanning a smooth distribution in the sense of  Stefan--Sussmann, with the particularity that the initial LH system and its  Vessiot--Guldberg Lie algebra is retrieved when $z\to 0$. This allowed us, among other applications, to provide a unified geometrical description of  the PH deformations of the three inequivalent LH systems on the plane based on the Lie algebra $\mathfrak{sl}(2,\mathbb{R})$ (see~\cite{BB2015,BCFHLb}). Nonetheless, the PH deformation method proposed in~\cite{BCFHL} was, to a certain extent, still incomplete, as it did not studied the existence and methods of derivation of an extension of the superposition rule concept for PH deformed LH systems, a hereafter called {\it PH deformed superposition rule} or, simply, a {\it deformed superposition rule}. 
  
  In this paper we present the way to implement such a generic procedure, by making use of a powerful tool developed in the context of superintegrable systems possessing  a Hopf algebra symmetry~\cite{BBHMO2009}. The construction is proved to be complete and valid for the hereafter called  prolonged PH deformations of LH systems, hence providing a generic prescription for the obtention of their deformed superposition principles. In a nutshell, deformed superposition rules are $z$-parametric families of mappings allowing for the description of some coordinates of particular solutions of a prolonged PH deformation of a LH system in terms of the others. When $z\rightarrow 0$, deformed superposition rules and prolonged PH deformations become standard superposition rules and diagonal prolongations \cite{LSbook}  for LH systems, respectively. It is remarkable that given a LH system on an $n$-dimensional manifold $M$, its prolonged PH deformations become Hamiltonian systems on $M^{m+1}$ that neither need to be Lie systems nor must consist of several copies of the original Lie system for $z\neq 0$. This is a remarkable difference with respect to LH systems, whose diagonal prolongations to $M^{m+1}$ give rise to a LH system consisting of several copies of the initial one and whose constants of the motion, for $m$ large enough, allow for a superposition rule for the initial LH system.

The paper is structured as follows. In Section \ref{s2} the fundamental properties of LH systems and their deformations based on the notion of PH algebras, as developed in~\cite{BCFHL}, are shortly reviewed. Special attention is devoted to the construction of constants of the motion of both LH systems and their prolonged PH deformations, which are hereafter called {\it prolonged deformations} to simplify our terminology. While in Subsection~\ref{s231} we recall  the method deduced from the coalgebra formalism~\cite{B2013}, already used in~\cite{BCFHL} and~\cite{BCFHLb}, in Subsection~\ref{s232} we present new material that completes and enlarges the previous work. Specifically, it is shown that for generic prolonged deformations of LH systems, in contrast to what happens for the undeformed systems, constructing the constants of the motion basing solely on the coalgebra formalism~\cite{B2013}  may not supply us with a sufficient number of functionally  independent  constants to establish a deformed superposition rule in explicit closed form, as in this case we necessarily have to consider constants of the motion of higher `order' (i.e.~the dimension of the underlying tensor product space for the coproduct), which could imply that the resulting expressions are not expressible analytically in a discernible way. This rather subtle phenomenon is due to  a symmetry breaking phenomena in the  coproduct induced by the deformation, which also causes that the prolonged deformation of the initial LH system is not, in general, a diagonal prolongation \cite{LSbook}.  In turn, this implies that a permutation in the tensor product space does not necessarily transform  a constant of the motion for the prolonged deformation into another one.   Moreover, it can even happen that some of the constants of the motion of the diagonal prolongation of an undeformed LH system do not have at all a counterpart for a prolonged deformation of the LH system. This can be seen as a gap in the formalism as given in~\cite{BCFHL}, as it somewhat disturbs the correspondence between the diagonal prolongations of a LH system and its prolonged deformations through the limit with respect to the deformation parameter. Based on this observation, we reconsider the problem of determining a sufficient number of constants of the motion for both the diagonal prolongation of a LH system and its prolonged  deformations, in such a manner that the correspondence through the limit $z\to 0$ is maintained, and we satisfactorily solve it in full generality,  using a   mechanism borrowed from the theory of superintegrable systems, namely using the superintegrability property of systems endowed  with a Hopf algebra of symmetries~\cite{BBHMO2009}. This construction leads to the consideration of a left- and a right-set of  $m$  constants of the motion in involution (for each set) that are valid for arbitrary prolonged deformations to $M^{m+1}$, providing a      maximal number of   $(2m-1)$ functionally independent  constants of the motion  for the given deformation.  From them we can infer a suitable deformed superposition rule selecting those constants of the motion having minimal `order', hence minimizing the analytical computation difficulties. In the classical limit $z\to 0$, the  `additional' right-constants of the motion are shown to be obtainable using the permutation method from the left-set, which was the one formerly considered in~\cite{B2013}. This new Ansatz refines in a natural way the results in~\cite{BCFHL}, pointing out the relevance of working simultaneously with left- and right-coproducts in the deformation, as it constitutes a procedure that generically guarantees that the resulting functions are constants of the motion of the prolonged deformation of the initial LH system.

  As an   application, Section \ref{s3} analyses the LH systems based on the oscillator algebra $\mathfrak{h}_4$. The (undeformed) diagonal prolongations to $\mathfrak{h}_4$-LH systems, which were already studied in \cite{Ba2015}, are now considered by using a different set of constants of the motion, leading to a different albeit equivalent superposition rule. The purpose  of this reformulation is to consistently deduce the superposition rule as the limit $z\to 0$ of the deformed one, whose derivation involves the use of the right-set of constants of the motion. 
  In Section \ref{s4}, the so-called  nonstandard (or Jordanian) PH deformation  of $\mathfrak{h}_4$~\cite{h4}  is considered
  and the explicit derivation of formulae for the corresponding prolonged deformation of  $\mathfrak{h}_4$-LH systems is developed. 
The choice of the nonstandard   deformation of $\mathfrak{h}_4$, on the other hand, has an interesting structural consequence: the two-dimensional book Lie algebra $\mathfrak{b}_2$ is preserved as a Hopf subalgebra, thus allowing us to restrict the  prolonged deformation of $\mathfrak{h}_4$-LH systems to this subalgebra and therefore to obtain deformed $\mathfrak{b}_2$-LH systems. As a representative of such systems based on $\mathfrak{b}_2$, we consider the complex Bernoulli equations, for which the constants of the motion and the superposition rules are given for both deformed and non-deformed versions. In this context, the interpretation of the deformed system as a small perturbation of the initial one, governed by the deformation parameter $z$, allows us to establish a connection between non-trivially coupled systems, and one of its equations corresponds to a Riccati equation. Finally, in Section \ref{CR} some conclusions are drawn, several possible future developments of the method are proposed,  and its applications  to the analysis of nonlinear systems of differential equations are commented.


\section{The Poisson--Hopf   deformation formalism revisited}
\label{s2}

This section briefly recalls the fundamental properties of Lie and  LH  systems. It also reviews the 
general deformation method of LH systems introduced in~\cite{BCFHL} (see also~\cite{BCFHLb})    based on PH algebras. In contrast to \cite{BCFHL,BCFHLb}, the deformation of LH systems is developed in full generality to point out that the mechanism holds for arbitrary smooth manifolds. In particular, we focus on the problem of obtaining a sufficient number of constants of the motion for prolonged deformations of LH systems, from which formally   explicit deformed superposition rules can be deduced, regardless of the initial LH system and the considered deformation. It should be emphasized that, albeit in~\cite{BCFHL} this possibility was outlined for deformed LH systems, no deformed superposition rules were explicitly given. In this work we propose an extension of the techniques in~\cite{BCFHL}  that allows us to construct such deformed superposition rules for arbitrary prolonged deformations of LH systems. Unless otherwise stated, we hereafter assume all structures to be smooth and globally defined. This will simplify the presentation of our results and it will allows us to focus on its main new features.


\subsection{Lie and Lie--Hamilton systems}
\label{s21}

Let $\>x=\left\{x_1,\dots, x_n\right\}$ denote the coordinates in an $n$-dimensional manifold $\mm$  and consider a non-autonomous system of
first-order ordinary differential equations given  by
\begin{equation} 
\frac{{\rm d} x_j}{{\rm d}t}=f_j(t,\>x ), \qquad j=1,\dots, n,
 \label{aa}
\end{equation}
for some arbitrary functions $f_j:\mathbb{R }\times \mm\rightarrow \mathbb{R}$. Such a system can be described equivalently via a $t$-dependent vector field  ${\bf
X}:\mathbb{R}\times \mm\rightarrow {\rm T}\mm$ defined as:
\be
{\bf X} (t,\>x)=
\sum_{j=1}^{n} f_j(t, \>x)\frac{\partial}{\partial
x_j}.
 \label{ab}
 \ee
A system of the type (\ref{aa}) is called a {\em Lie system}~\cite{LS,VES,DAV,PW,CGM00,CGM07,CGL2010,CL2011,CGL12,LSbook} whenever it admits a {\it fundamental system of solutions}, i.e., whenever its general solution, ${\bf x}(t)$, can be expressed in terms of $m$  particular solutions
$\left\{  \mathbf{x}_{1}(t),\dots ,\mathbf{x}_{m}(t)\right\}  $ and $n$
constants $\left\{
k_{1},\dots,k_{n}\right\}  $ as
\be
{\bf x}(t)=\Psi (  \mathbf{x}_{1}(t),\ldots,\mathbf{x}_{m}(t),k_{1}
,\ldots,k_{n} )
 \label{ac}
 \ee
for a certain function $\Psi:\mm^{m}\times \mm\rightarrow \mm$. The expression (\ref{ac}) is  usually referred to as a {\em superposition rule} of 
the system (\ref{aa}).  Within applications to physical phenomena, the wealth of group-theoretical techniques developed from the 60's onwards revived the interest in analyzing systematically the existence of  superposition rules, leading to an extensive geometrical study of Lie systems and superposition rules and their application to systems at both the classical and quantum levels (see, e.g., \cite{LSbook, Insel1,Insel2,Levin,Kon,Shn,Joa,Doro}  and references therein).

The {Lie--Scheffers Theorem} (see~\cite{LS,VES,CGM00,LSbook}) states that a $t$-dependent vector field $\mathbf{X}$ as in (\ref{ab}) determines a Lie system if and only if there exist some functions $b_1(t),\ldots,b_\ell(t)$ and vector fields ${\bf X}_1,\ldots,{\bf X}_\ell$ on $\mm$
spanning an $\ell$-dimensional real Lie algebra $V$ 
such that 
\be
{\bf X}(t,\>x)=\sum_{i=1}^\ell b_i(t){\mathbf X}_i(\>x),\qquad \forall {\bf x}\in M.
\label{aabb} 
\ee
It can then be proved that the system $\> X$  admits a superposition rule so that the constraint $\ell\leq nm$ is satisfied. In these conditions, $V$ is called a {\em Vessiot--Guldberg Lie algebra} of ${\bf X}$ (see also \cite{CS2016,CS2016b,Ibragimov2016} for more recent applications of  Vessiot--Guldberg Lie algebras). 

A Lie system is said to be a {\em Lie--Hamilton system} whenever it admits a Vessiot--Guldberg Lie algebra $V$ of Hamiltonian vector fields   with respect to a Poisson structure  \cite{CL2011,LSbook,CLS2013,B2013,BB2015,Ba2015,HLT}. Let us assume the case of a LH system on $M$ that admits a Vessiot--Guldberg Lie algebra, $V$, of Hamiltonian vector fields relative to a symplectic form $\omega$. The compatibility condition between the generators ${\bf X}_i$ of  $V$ and $\omega$ is locally determined by the invariance of $\omega$ under the Lie derivative with respect to any generator $\>X_i$ of $V$, i.e.,
\be
{\cal L}_{\mathbf{X}_i}\omega =0, \qquad i=1,\dots, \ell.
\label{ad}
\ee
Now a {\em Hamiltonian function}  $h_i$ is related to the vector field ${\bf X}_i$ through the contraction or inner product of $\omega$ with respect to ${\bf X}_i$:
\be
\iota_{{\bf X}_i}\omega={\rm d}h_i,  \qquad i=1,\dots, \ell.
\label{ae}
\ee
Recall that every symplectic form allows us to define a  Poisson bracket 
\be
\{\cdot,\cdot\}_\omega\ :\   (f_1,f_2) \in C^\infty (\mm )\times C^\infty (\mm )\rightarrow {\bf X}_{f_2} f_1 \in C^\infty (\mm ),
\label{af}
\ee
where $\> X_{f}$ is the unique Hamiltonian vector such that $\iota_{\>X_f}\omega={\rm d}f$ for an $f\in C^\infty(M)$. It follows that $(C^\infty(\mm),\{\cdot,\cdot\}_\omega)$ is endowed with a Poisson algebra structure. The space ${\rm Ham}(\omega)$ of  Hamiltonian vector fields on $\mm$ relative to $\omega$, which is a Lie algebra with respect to the commutator of vector fields, is related to the former by means of the Lie algebra morphism~\cite{Vaisman}  
\be
 (C^\infty(\mm),\{\cdot,\cdot\}_\omega)\stackrel{\varphi}{\longrightarrow} ({\rm Ham}(\omega),[\cdot,\cdot])
 \label{ag}
 \ee
mapping a function $f\in C^\infty(\mm)$ onto the Hamiltonian vector field $-{\bf X}_f$.  The Hamiltonian functions $h_i$ $(i=1,\dots,\ell)$ coming from  (\ref{ae}) 
span, eventually together with a constant function $h_0$ on $\mm$, a finite-dimensional Lie algebra of functions $\varphi^{-1}(V)$ that is called a {\it Lie--Hamilton algebra} (LH algebra), ${\cal H}_\omega$, of $\>X$~\cite{BB2015, Ba2015}.


\subsection{The Poisson--Hopf deformation approach}
\label{s22}

The remarkable point is that the space $C^\infty({\cal H}_\omega^\ast)$  of smooth functions on the dual $\mathcal{H}^*_\omega$ of the LH algebra $\mathcal{H}_\omega$ can be endowed with a Hopf algebra structure~\cite{CP,majid,Abe}.
For our purposes, it suffices to consider the {\em coalgebra structure} of the Hopf algebra determined by the coproduct map, as the remaining structural maps, namely the counit and antipode, can be deduced from the axioms defining the Hopf algebra. In particular, for an associative algebra $A$, the coproduct $\Delta: A\to A\otimes A$  must be an algebra 
homomorphism and satisfy the coassociativity condition
\be
({\rm Id}\otimes\Delta)\Delta(a)=(\Delta\otimes {\rm Id})\Delta(a),\qquad \forall a\in A .
\label{ba}
\ee
If  $A$ is a commutative Poisson algebra, the coproduct $\Delta$ satisfying (\ref{ba}) is required to be a Poisson algebra morphism, so that the   Poisson bracket on $A\otimes A$ is given by
\be
\{ a\otimes b, c\otimes d\}_{A\otimes A}=\{ a, c\}\otimes  b d + a c\otimes \{ b, d\} ,\qquad \forall a,b,c,d\in A .
\label{bb}
\ee
For the case of $C^\infty({\cal H}_\omega^\ast)$, the coalgebra structure is determined by the coproduct  $ \Delta (f)(x_1,x_2):=f(x_1+x_2)$, where  $x_1,x_2\in {\cal H}_\omega$ and $f\in C^\infty({\cal H}_\omega^\ast)$. The details concerning the complete Hopf algebra structure can be found in~\cite{BCFHL}; here we just recall that $C^\infty\left(\lh^* \right)$ turns out to be a PH algebra through the Poisson structure defined by the Kirillov--Kostant--Souriau bracket related to a Lie algebra structure on ${\cal H}_\omega$. 

Given these algebraic preliminaries, we summarize the notion of PH deformation introduced in~\cite{BCFHL} (see also~\cite{BCFHLb}) in four steps: 
 
\begin{enumerate}

\item Let ${\bf X}$ be a LH system  of type (\ref{aabb})  on  an $n$-dimensional manifold $\mm$ with symplectic form $\omega$, so that the LH algebra $\mathcal{H}_{\omega}$ is spanned by a set of functions $\left\{h_1,\ldots, h_\ell\right\}\subset C^\infty(\mmm)$ satisfying the condition  (\ref{ae}),  with $\mmm$ being a suitable  submanifold of $\mm$ that ensures that each $h_i$ is well defined.  Let the Poisson bracket of the functions $h_i$ be given by:
\be
\{h_i,h_j\}_{\omega}= \sum_{{\sss=1}}^{\ell} C_{ij}^{\sss}h_{\sss},\qquad  i,j=1,\dots, \ell,
\label{bd}
\ee
for certain structure  constants $C_{ij}^\sss$. 
\item Consider a PH deformation of  $C^\infty({\cal H}_\omega^\ast)$, denoted by $\qlhz$, with   deformation parameter $z\in\mathbb R$ (or $q={\rm e}^z$) as the space of smooth functions $F_{z,ij}(h_{z,1},\ldots,h_{z,\ell})$ for a family of functions $\{ h_{z,1},\ldots,h_{z,\ell} \}$  on $ C^\infty(\mmm)$ with Poisson bracket (with respect to $\omega$) given by
\be
\{h_{z,i},h_{z,j}\}_{\omega}= F_{z,ij}(h_{z,1},\dots,h_{z,\ell}),\qquad  i,j=1,\dots, \ell,
\label{be}
\ee
and satisfying the non-deformed limits 
\be
  \lim_{z\to 0} h_{z,i}=h_i ,\qquad   \lim_{z\to 0}   F_{z,ij}(h_{z,1},\dots,h_{z,\ell }) = \sum_{{\sss=1}}^{\ell} C_{ij}^\sss h_\sss   .
\label{bf}
\ee

\item Obtain the deformed vector fields ${\bf X}_{z,i}$   according to  the relation (\ref{ae}), that is,
\be
\iota_{{\bf X}_{z,i}}\omega={\rm d}h_{z,i},  \qquad i=1,\dots, \ell .
\label{bg}
\ee

\item And, finally, define the PH deformation ${\bf X}_z$ of the LH system ${\bf X}$   (\ref{aabb})  as 
\be
{\bf X}_{z}:=\sum_{i=1}^{\ell} b_i(t){\bf X}_{z,i}.
\label{bh}
\ee
\end{enumerate} 
Notice that, by construction, the following non-deformed limits are consistently recovered:
\begin{equation}
\lim_{z\to 0} {\bf X}_{z,i}= {\bf X}_i,\qquad \lim_{z\to 0} {\bf X}_{z}={\bf X} .
\label{bi}
\end{equation}
The essential point to be taken into account is that the deformed vector fields $\{ {\bf X}_{z,1},\dots,  {\bf X}_{z,\ell} \}$ obtained through the preceding prescription do not, in general,  provide neither a finite-dimensional Lie algebra nor a quantum algebra. Actually, they span an involutive Stefan--Sussmann distribution~\cite{Vaisman,Pa57,WA} since
\be
[{\bf X}_{z,i},{\bf X}_{z,j}]= -\sum_{\sss=1}^\ell\frac{\partial F_{z,ij}}{\partial h_{z,\sss}}\, {\bf X}_{z,\sss} ,\qquad  i,j=1,\dots, \ell.
\label{bj}
\ee
In other words, the functions $\{ h_{z,1},\ldots,h_{z,\ell} \}$ determine a PH deformation $\qlhz$ of $C^\infty({\cal H}_\omega^\ast)$ with deformed Poisson brackets (\ref{be}), thus providing a (deformed) Hamiltonian function
\be
h_{z}:=\sum_{i=1}^\ell b_i(t)h_{z,i}.
\label{bk}
\ee
However, the non-autonomous system of
first-order ordinary differential equations ${\bf X}_{z}$   (\ref{bh}) does no longer correspond, in general, to a Lie system, but to a `perturbation'  of the initial system (\ref{aa}) with respect to the   deformation parameter $z$, as follows at once from the conditions (\ref{bf}) and (\ref{bi}) 
under the limit $z\to 0$. In this context, it is conceivable to interpret $z$  as a small perturbation parameter. This means that, once the deformed system has been obtained through either ${\bf X}_{z}$ or $h_{z}$, a power series expansion in $z$  can be considered, analyzing the behaviour of the deformed Hamiltonian system up to the first, second or some higher order, enabling us a comparison with the initial undeformed system.


\subsection{Constants of the motion}
\label{s23}

The coalgebra formalism considered in~\cite{BCR96,BR98} in the context of integrable systems turned out to be a highly effective tool that allows to prove in a constructive way the complete integrability of systems possessing coalgebra symmetry, including the explicit construction of the corresponding integrals of the motion. This coalgebra approach was later extended in order to characterize the property of (quasi-maximal) superintegrability~\cite{BHMO2004,BH2007,BBHMO2009}. These results covered both non-deformed integrable systems and their PH deformations. More recently, the coalgebra formalism was adapted to the framework of LH systems~\cite{B2013}, providing a method to determine $t$-independent constants of the motion in a more direct way than that given by the classical methods \cite{PW,CGM07}. We observe that the constants of the motion of LH systems deduced by this technique are the cornerstone for the obtention of superposition rules. This procedure has been carried out systematically for LH systems on the plane ${\mathbb R}^2$ in~\cite{Ba2015} as well as on two-dimensional spaces of constant curvature  (with different signatures of the metric tensor) in~\cite{HLT}. 

At this point, it is of capital importance to realize that the results presented in~\cite{B2013} (and further considered in~\cite{Ba2015,HLT}) focused on non-deformed LH systems, that is, for cases with a trivial or primitive coalgebra structure. This approach turns out be unsatisfactory, as its straightforward extension to non-primitive coproducts,  which is precisely the case for  prolonged deformations of LH systems, provides less  constants of the motion than in the primitive case. The aim of this section is to enlarge and complete such previous work, proposing a general procedure for the explicit construction of the  constants of the motion for prolonged   deformations  of LH systems.


\subsubsection{Undeformed constants of the motion}
\label{s231}
 
Let us first briefly summarize the coalgebra  approach for constructing $t$-independent constants of the motion of non-deformed LH systems~\cite{B2013} (see also~\cite{BCFHL,Ba2015}). 
Consider the  LH algebra ${\cal H}_\omega$ of a LH system  $\>X$~(\ref{aabb}), expressed as a Lie--Poisson algebra with generators $\{ v_1,\dots,v_\ell \}$ fulfilling the Poisson brackets (see (\ref{bd})):
 \be
\{v_i,v_j\} = \sum_{{\sss=1}}^{\ell} C_{ij}^{\sss}v_{\sss},\qquad  i,j=1,\dots, \ell.
\label{ca}
\ee
Let   $S\left(\lh\right)$ be the   symmetric algebra  of $\lh$ (i.e., the associative unital algebra of polynomials in the elements of $\lh$)  understood as a Poisson algebra, thus with fundamental Poisson brackets (\ref{ca}). Under these conditions, $S\left(\lh\right)$  can always be  endowed with a   coalgebra structure with a  non-deformed (trivial or primitive) coproduct map $\Delta$ defined  by
\begin{equation}
 {\Delta} :S\left(\lh \right)\rightarrow
S\left(\lh\right) \otimes S\left(\lh\right)    ,\qquad      {\Delta}(v_i):=v_i\otimes 1+1\otimes v_i,  \qquad    i=1,\dots, \ell,
\label{cb}
\end{equation}
which is a Poisson algebra homomorphism of (\ref{ca}). Notice that  the (trivial) counit and   antipode can also be defined giving rise to the non-deformed Hopf structure corresponding to any Lie algebra~\cite{CP,majid,Abe}.

 The 2-coproduct $\Delta\equiv \Delta^{(2)}$ can be extended to a third-order coproduct through  the coassociativity condition (\ref{ba}):
\bea
&&\Delta^{(3)}:=(\Delta \otimes {\rm Id}) \circ \Delta=({\rm Id} \otimes \Delta) \circ \Delta ,
 \nonumber\\[2pt]
&&
\Delta^{(3)}:\  S(\lh) \rightarrow
S(\lh) \otimes S(\lh)\otimes S(\lh)\equiv   S^{(3)}(\lh) ,
\label{cc}\\[2pt]
&&{\Delta}^{(3)}(v_i)=v_i\otimes 1\otimes 1 +1\otimes v_i\otimes 1+1\otimes 1\otimes v_i ,\qquad i=1,\ldots,\ell.
\nonumber
\eea
A  $k^{th}$-order coproduct map  can be defined recursively by the rule 
\bea
&& \Delta ^{(k)}: \ S(\lh) \rightarrow {\stackrel{ k\ {\rm times} } {\overbrace{   S(\lh) \otimes\ldots\otimes   S(\lh)  }}  }\equiv  S^{(k)}(\lh) , \nonumber\\[2pt]
&& {\Delta}^{(k)}:= \bigr({\stackrel{(k-2)\ {\rm times}}{\overbrace{{\rm
Id}\otimes\ldots\otimes{\rm Id}}}}    \otimes {\Delta^{(2)}}  \bigr)\circ \Delta^{(k-1)},
\label{cd}
\eea
 which  is also  a Poisson algebra homomorphism for any $k\ge 3$.

Any element  of $S(\lh)$ can be seen as a function on $\lh^*$ so that the coproduct  (\ref{cb}) in $S(\mathcal{H}_\omega)$ can be extended  to
\begin{equation}
 {\Delta} :C^\infty\left(\lh^*\right)\rightarrow
C^\infty\left(\lh^*\right) \otimes C^\infty\left(\lh^*\right)\subset C^\infty(\lh^*\times \lh^*).
\label{ce}
\end{equation}
A similar extension holds for the $k^{th}$-order coproduct defined in (\ref{cd}).  Therefore,  $C^\infty(\mathcal{H}^*_\omega)$ becomes a non-deformed Poisson coalgebra, and the corresponding extension of the counit and antipode maps turns  $C^\infty(\mathcal{H}^*_\omega)$ into a PH algebra \cite{BCFHL}. 

Now consider the LH algebra $\lh$ spanned by the   Hamiltonian functions  $\{ h_1,\dots,h_\ell \}$ satisfying the Poisson brackets (\ref{bd}). In agreement with equation (\ref{ca}), we define  the Lie algebra morphism  
\be
\phi:\lh\rightarrow C^\infty(\mmm),\qquad h_i:=\phi(v_i),\qquad i=1,\ldots,\ell,
\label{cf}
\ee 
where $\mmm\subset \mm$  is chosen in order to ensure that the functions $h_i$, and their PH deformations, to be defined shortly, are well defined. Basing on this result, we construct a family of Poisson algebra morphisms 
\be
 D: C^\infty\left( \lh^* \right) \rightarrow C^\infty(\mmm),\,\,\   D^{(\kk)} :     {\stackrel{ \kk\ {\rm times} } {\overbrace{   C^\infty\left(   \lh^* \right)\otimes \ldots\otimes C^\infty\left(  \lh^* \right) }}}   \rightarrow  {\stackrel{ \kk\ {\rm times} } {\overbrace{ C^\infty(\mmm)\otimes\ldots\otimes C^\infty(\mmm) }}}\subset C^\infty(\mathcal{M}^k),
\label{cg}
\ee
that are given by   
\be
D( v_i)= h_i(\>x_1) := h^{(1)}_i , \qquad
 D^{(\kk)} \bigl( {\Delta}^{(k)}(v_i) \bigr)= h_i(\>x_1)+\cdots + h_i(\>x_\kk)  := h^{(\kk)}_i ,\qquad i=1,\dots, \ell,
\label{ch}
\ee
where  $\>x_j=\left\{(x_1)_j,\dots, (x_n)_j\right\}$ denotes the coordinates in the  $j$-copy submanifold $\mmm\subset\mm$ within $\mathcal{M}^k$.  In the generic case with a $k^{th}$-order    tensor product of elements $u_j(v_1,\dots,v_\ell)\in  C^\infty\left(   \lh^* \right)$ the morphism $ D^{(\kk)}$ gives rise to the following product of functions
\be
 D^{(\kk)} \bigl( u_1(v_1,\dots,v_\ell) \otimes \ldots \otimes u_k(v_1,\dots,v_\ell)   \bigr)= u_1(h_1(\>x_1) ,\dots,h_\ell(\>x_1) ) \dots
  u_k(h_1(\>x_k) ,\dots,h_\ell(\>x_k) ).
 \label{chx}
\ee

Let us finally assume that   $C^\infty\left(\lh^*\right)$  
possesses a Casimir  invariant 
\be
C=C(v_1,\dots,v_\ell),
\label{ci}
\ee
that is, an element $C$ that Poisson-commutes with all $v_i$ with respect to the Poisson bracket given in  (\ref{ca}). As proved in~\cite{B2013}, it follows that the functions constructed through the family of coproducts (\ref{cd}) and Poisson morphisms (\ref{cg})  defined by
\begin{equation}
  F:= D(C),\qquad F^{(\kk)}\left(h^{(\kk)}_1,\dots,h^{(\kk)}_\ell\right) := D^{(\kk)}\left[\Delta^{(\kk)} \bigl({C( v_1,\dots,v_\ell)} \bigr) \right] ,   \qquad \kk=2,\ldots,m+1,
  \label{cj}
\end{equation}
are  $t$-independent constants of the motion for the diagonal prolongation
$\widetilde {\>X}^{m+1}$  of the LH system $\>X$  (\ref{aabb})  to the product manifold $\mmm^{m+1}$, i.e., the $t$-dependent vector field on $\mathcal{M}^{m+1}$ of the form
\begin{equation}\label{DP}
\widetilde{\>X}^{m+1}(t,\>x_1,\ldots,\>x_{m+1}):=
\sum_{k=1}^{m+1}\sum_{j=1}^nX^j(t,\>x_k)\frac{\partial }{\partial x^j}=\sum_{i=1}^\ell b_i(t)\>X_{ h^{(m+1)}_i} \, .
\end{equation}
Note that the functions (\ref{cj}) can also be considered as constants of the motion for the  LH system $\>X$. 
The right-hand side of expression (\ref{DP}) shall be called the {\it prolonged PH deformation}  of $\bf X$ to $\mathcal{M}^{m+1}$, or simply the {\it prolonged deformation}. As we shall see shortly, this notion can immediately  be  extended to non-primitive coproducts which will invalidate, in general, the equality of the right-hand side of (\ref{DP}) with the standard diagonal prolongation of $\> X$. 

Each of the $F^{(k)}$  (\ref{cj}) can be considered as a function of $C^\infty(\mmm^{m+1})$.  lf all the $F^{(k)}$ are non-constant, then  they form a set of  $m$  functionally independent functions in $C^\infty(\mmm^{m+1})$ that are in involution. In addition, these functions $F^{(k)}$ can be used to generate other 
$t$-independent  constants of the motion   by means of the prescription~\cite{B2013}
\begin{equation}
F_{ij}^{(k)}=S_{ij} \bigl( F^{(k)}   \bigr) , \qquad 1\le  i<j\le  k,\qquad k=2,\ldots,m+1,
\label{ck}
\end{equation}
where $S_{ij}$ denotes the permutation of the variables $\>x_i\leftrightarrow
\>x_j$ in $\mmm^{m+1}$. This can be viewed as a consequence of  the fact that  the diagonal prolongation of $\> X$ is invariant under such a permutation of variables. It can also be viewed as a consequence of (\ref{DP}) and \cite[Proposition 1]{BR98}.

Recall that for obtaining a superposition rule depending on $m$ particular solutions, as in  (\ref{ac}), one searches for a set, $I_1,\ldots,I_n$, of $t$-independent constants of the motion on $\mathcal{M}^{m+1}$ for $\widetilde{\> X}^{m+1}$ so that \cite{LSbook}
\be
\frac{\partial (I_1,\ldots,I_n) }{\partial ( (x_1)_{m+1},\dots, (x_n)_{m+1}   )}\neq 0,
\ee
and the diagonal prolongations $\widetilde{\>X}^m_1,\ldots,\widetilde{\>X}^m_\ell$ are linearly independent at a generic point \cite{LSbook}. This allows us to express the coordinates  $\>x_{m+1}=\left\{ (x_1)_{m+1},\dots, (x_n)_{m+1}\right\}$     in terms of the remaining coordinates in $\mathcal{M}^{m+1}$ and the constants $k_1,\ldots,k_n$   defined by  the conditions $I_1=k_1,\ldots,I_n=k_n$.
We stress that the set of  constants of the motion  (\ref{cj}) and  (\ref{ck}) are frequently sufficient to deduce the superposition rules for the LH system $\>X$   (\ref{aabb}) in a direct way, as it has already been explicitly shown in~\cite{Ba2015,HLT} for some specific LH systems. Moreover, the existence of   a large number of constants of the motion $F_{ij}^{(k)}$ (obtained through permutations) rather simplifies the computations, as it allows one to keep the number $k$   low. 
   Actually, in most of the explicit superposition rules worked out in~\cite{Ba2015,HLT}, it was sufficient to consider the function $F^{(2)}$ and its permutations $F_{ij}^{(2)}$, a fact that helped to avoid long cumbersome computations enabling to establish in closed form a superposition principle of reasonable simplicity. However, as we shall prove in the sequel, the functions (\ref{cj}) and  (\ref{ck}) will not generically provide us, in the case of  prolonged   deformations of a LH system, with a sufficient number of functionally independent constants of the motion from which a deformed superposition rule could easily be  inferred.


\subsubsection{Deformed constants of the motion}
\label{s232}

As it has been already stated, the fact that $ \qlhzd$ is a PH    deformation of $C^\infty({\cal H}_\omega^\ast)$ enables us  to apply  the coalgebra formalism proposed in~\cite{B2013} to construct  $t$-independent constants of   motion for the deformed LH system ${\bf X}_z$   (\ref{bh}) with  the deformed Hamiltonian $h_z$ as given in (\ref{bk}),  for which some examples were presented in~\cite{BCFHL}. This procedure must however be applied with some caution, as there are some subtle points that, if not taken into account, may invalidate the conclusions. The key point is to observe that, whenever we are considering  a deformed PH algebra, the deformed  coproduct $\Delta_z$ is no longer trivial (or primitive) as $\Delta$ (\ref{cb})  for all the generators $v_i$. Indeed, the deformation `breaks' the symmetry within the coproduct, that is, the positions in the tensor product space within the coproduct are no longer `equivalent', as happens e.g.~for~${\Delta}^{(3)}(v_i)$ in (\ref{cc}).

By the construction in~\cite{BCR96,BR98}, the deformed counterpart of the constants of the motion $F^{(k)}$ (\ref{cj}) still holds, but it is not ensured that the permutations in (\ref{ck}) give rise to $t$-independent constants of the motion for the prolonged   deformation   $\widetilde{\> X}^{m+1}_z$ to $\mathcal{M}^{m+1}$ of ${\>X}$.
Therefore, in the deformed case one may need to consider a higher number $k$ with respect to the non-deformed system to deduce the deformed superposition rule, which in turns makes $m+1$ to be larger. The drawback of considering an increased $m+1$ is that the complexity of the computations grows exponentially, resulting in an extremely involved derivation of the deformed superposition rule. Fortunately, this difficulty can be circumvented by considering a second set of constants of the motion, additionally to the $F^{(\kk)}$, which comes from the superintegrability property of integrable systems possessing Hopf algebra symmetry~\cite{BHMO2004,BH2007,BBHMO2009}. In the following we present the explicit derivation of both sets of deformed constants of the motion.

    Let ${\cal H}_{z,\omega}$ be the    deformed   LH algebra of the deformed LH system  $\>X_z$  (\ref{bh})  with   Hamiltonian $h_z$ given by (\ref{bk}). We take a  basis with generators $\{ v_1,\dots,v_\ell \}$  such that the Poisson brackets are given by (see (\ref{be})):
\be
\{v_{i},v_{j}\}_{z}= F_{z,ij}(v_{1},\dots,v_{\ell}),\qquad  i,j=1,\dots, \ell.
\label{da}
\ee
Proceeding as in the non-deformed case, we consider the deformed coproduct for the generators $v_i$:
\begin{equation}
 {\Delta}_z : \qlhzd\rightarrow
\qlhzd \otimes \qlhzd,
\label{db}
\end{equation}
along with the  $k^{th}$-order deformed coproduct map, ${\Delta}_z^{(k)}$, defined exactly as in (\ref{cd}), such that the limits
\be
\lim_{z\to 0} \Delta_{z}=\Delta ,\qquad \lim_{z\to 0} \Delta_{z}^{(k)}=\Delta^{(k)} ,
\ee 
are satisfied.

Now recall that the deformed Hamiltonian functions $\{ h_{z,1},\ldots,h_{z,\ell} \}$ fulfil the relations (\ref{bg}) and  the      Poisson brackets   (\ref{be})  with respect to the symplectic form $\omega$.  We define  the map
\be
\phi_z: {\cal H}_{z,\omega}\rightarrow  C^\infty(\mmm),\qquad h_{z,i}:=\phi_z(v_i),\qquad i=1,\ldots,\ell,
\label{dc}
\ee 
where again $\mmm\subset \mm$  is chosen to guarantee that the functions $h_{z,i}$ are properly defined. Next, as in (\ref{cg}),  we introduce  the Poisson algebra morphisms 
\be
 D_z: \qlhzd \rightarrow C^\infty(\mmm),\quad\   D_z^{(\kk)} :     {\stackrel{ \kk\ {\rm times} } {\overbrace{   \qlhzd \otimes \ldots\otimes \qlhzd }}}   \rightarrow  {\stackrel{ \kk\ {\rm times} } {\overbrace{ C^\infty(\mmm)\otimes\ldots\otimes C^\infty(\mmm) }}} .
\label{dd}
\ee
 Now let 
\be
C_z=C_z(v_1,\dots,v_\ell),
\label{de}
\ee
be the  Casimir function of $ \qlhzd$ with  $ \lim_{z\to 0} C_{z}= C$. And we define the functions (see (\ref{ch}))
\be
D_z( v_i) = h_{z,i} :=    h^{(1)}_{z,i}  , \qquad D_z^{(k)} \bigl(\Delta_z^{(k)}(v_i) \bigr)  :=  h^{(k)}_{z,i} ,\qquad i=1,\ldots,\ell  ,
\label{de2}
\ee
whose  explicit form  does depend  on the  initial deformed  coproduct  ${\Delta}_z$.  Anyhow, the analogous relation to (\ref{chx}) also holds:
\be
 D_z^{(\kk)} \bigl( u_1(v_1,\dots,v_\ell) \otimes \ldots \otimes u_k(v_1,\dots,v_\ell)   \bigr)= u_1(h_{z,1}(\>x_1) ,\dots,h_{z,\ell}(\>x_1) ) \dots
  u_k(h_{z,1}(\>x_k) ,\dots,h_{z,\ell}(\>x_{k}) ),
 \label{chxy}
\ee
with $u_j\in \qlhzd$,   allowing one to compute (\ref{de2}) from a given ${\Delta}_z$.

 The   first set of  constants of the motion for the prolonged   deformation of the LH system $\> X$, which is analogous to the  right-hand side of (\ref{DP}), i.e. 
 \be
 \widetilde{\>X}^{m+1}_z=\sum_{i=1}^\ell b_i(t) \, \> X_{h^{(m+1)}_{z,i}},
 \label{de3}
 \ee
  is defined by
\begin{equation}
  F_z:= D_z(C_z),\qquad F_z^{(\kk)}\left(h^{(k)}_{z,1},\dots,h^{(k)}_{z,\ell}\right) := D_z^{(\kk)}\left[\Delta_z^{(\kk)} \bigl({C_z( v_1,\dots,v_\ell)} \bigr) \right] ,   \qquad \kk=2,\ldots,m+1,
  \label{df}
\end{equation}
 which is just the deformed counterpart of   (\ref{cj}). It is worth noting that  $\> X_{h^{(m+1)}_{z,i}}$ in (\ref{de3}) fulfil similar commutation relations to  (\ref{bj}):
\be
\left[\>X_{h^{(m+1)}_{z,i}}, \>X_{h^{(m+1)}_{z,j}}\right]=-\sum_{\sss=1}^{\ell}\frac{\partial F_{z,ij}}{\partial h^{(m+1)}_{z,\sss}}\left(h^{(m+1)}_{z,1},\ldots,h^{(m+1)}_{z,\ell} \right) \, \>X_{h^{(m+1)}_{z,\sss}},\qquad i,j=1,\ldots,\ell .
  \label{df3}
\ee

  If   all $F_z^{(k)}$  (\ref{df}) are non-constant functions,    they  provide a set of  $m$      functionally independent functions  in involution~\cite{B2013,BCR96,BR98}. Even if formally these invariants are sufficient to deduce a deformed superposition rule, it is doubtful that a closed analytical expression can be obtained, as 
 the difficulty of the formulae increases exponentially when augmenting the order of the constants of the motion.  The crucial difference with the undeformed case is that the validity of the permutation process (\ref{ck}) is not guaranteed any more, as a consequence of the `broken-symmetry' of the deformed  coproduct $\Delta_z$ in the tensor product space. Hence, the deformed prolongation $\widetilde{\> X}^{m+1}_z$ to $\mathcal{M}^{m+1}$ of ${\>X}$ is not, in general, invariant  relative to the interchange of variables $\>x_i\leftrightarrow\>x_j$. In fact, only under the non-deformed limit  $z\to 0$,   the coproduct  $\Delta_z$ becomes  primitive  and $\widetilde{\> X}^{m+1}_z$ (\ref{de3}) reduces to $\widetilde{\> X}^{m+1}$  (\ref{DP}), being the latter  symmetric under such permutations.
   Thus, in principle, no additional constants of the motion can be obtained with this Ansatz for a generically prolonged   deformation of a LH system.
Nevertheless, following the approach to superintegrability of integrable systems with   coalgebra symmetry~\cite{BHMO2004,BH2007,BBHMO2009}, we can construct a second set of constants of the motion that is valid for any deformed LH system.

The essential point is that the  $k^{th}$-order coproduct  ${\Delta}_z^{(k)}$ is defined  within the $(m+1)^{th}$-order  tensor product space in the form
\be
{\stackrel{ \kk\ {\rm times} } {\overbrace{   \qlhzd \otimes \ldots\otimes \qlhzd }}}   \,\otimes\! \!
 {\stackrel{ {(m+1-k)} \ {\rm times} } {\overbrace{1\otimes \ldots\otimes 1}}} ,
 \ee
 and as a shorthand notation for the space where this object is defined we use
\be
1\otimes 2\otimes
\ldots \otimes k .
\label{dg}
\ee
However, instead of using (\ref{cd}), it is possible to define another recursion relation for
the $k^{th}$-order coproduct, as done in~\cite{BHMO2004,BBHMO2009}: 
 \be
\Delta_{zR}^{(k)}:= \bigr(  {\Delta_z^{(2)}} \otimes  {\stackrel{(k-2)\ {\rm times}}{\overbrace{{\rm
Id}\otimes\ldots\otimes{\rm Id}}}}     \bigr)\circ  \Delta_{zR}^{(k-1)},\qquad k\ge 3.
\label{dh}
 \ee
 Since we are considering products in the reversal ordering, it follows that  $ \Delta_{zR}^{(k)}$ lives in  the $(m+1)^{th}$-order  tensor product space
 \be
 {\stackrel{ {(m+1-k)} \ {\rm times} } {\overbrace{1\otimes \ldots\otimes 1}}}  \! \!  \otimes\, {\stackrel{ \kk\ {\rm times} } {\overbrace{   \qlhzd \otimes \ldots\otimes \qlhzd }}}  
\,  ,
 \ee
which will be  shortened as
 \be
 (m-k+2)\otimes (m-k+3)\otimes \ldots\otimes (m+1).
 \label{di}
 \ee
 The maps  ${\Delta}_z^{(k)}$  and   ${\Delta}_{zR}^{(k)}$ are called left- and right-coproducts, respectively. For this reason, we call 
 (\ref{df}) the {\em set of left-constants of the motion} for the prolonged deformation $\widetilde{\> X}^{m+1}_z$ of the LH system, while the corresponding  {\em set of right-constants of the motion} is defined by
\begin{equation}
 F_{z{(\kk)}}\left(h^{(k)}_{zR,1},\dots,h^{(k)}_{zR,\ell}\right) := D_{zR}^{(\kk)}\left[\Delta_{zR}^{(\kk)} \bigl({C_z( v_1,\dots,v_\ell)} \bigr) \right] ,   \qquad \kk=2,\ldots,m+1,
  \label{dk}
\end{equation}
where the morphisms $D_{zR}^{(\kk)}$ are defined as in (\ref{dd}), but now on the  right-tensor product space (\ref{di}), in such  a manner  that the functions 
$h^{(k)}_{zR,i}$ are defined by
\be
D_{zR}^{(k)} \bigl(\Delta_{zR}^{(k)}(v_i) \bigr)  := h^{(k)}_{zR,i} ,\qquad i=1,\ldots,\ell,\qquad k=2,\ldots,m+1 .
\label{df2}
\ee

It is straightforward to verify that, due to the coassociativity property  (\ref{ba}), the identity $\Delta_{zR}^{(m+1)}\equiv \Delta_{z}^{(m+1)}$ holds~\cite{BHMO2004,BBHMO2009}, which imples that  $  F_{z{(m+1)}}\equiv  F_z^{(m+1)}$.    Again, if all the $F_{z(k)}$  are non-constant,    they   constitute  a set of  $m$  functionally independent functions  in involution.  We stress that functional independence among all the integrals follows, by construction,  from the different tensor product spaces on which  they are defined. Furthermore, the two sets $ F_z^{(k)}$ and $F_{z{(k)}}$ altogether provide  a  maximal number of   $(2m-1)$ functionally independent   constants of the motion which are  valid for arbitrary PH deformations. Focusing on those functions having the lowest value of $k$, a closed analytical expression for the deformed superposition rule can be much more easily found that merely considering the set of left-constants of the motion.

For completeness in the exposition, we display  the constants of the motion corresponding to both sets in Table~\ref{table1} using the shorthand notations (\ref{dg}) and (\ref{di}). Under the non-deformed limit $z\to 0$, the left-set $F_z^{(\kk)}$ (\ref{df}) reduces to $F^{(\kk)}$ (\ref{cj}), while 
the right-set $ F_{z{(\kk)}}$ (\ref{dk}) provides constants of the motion  $F_{{(\kk)}}$ for the undeformed LH system that are expressible in terms of the set of permutations  (\ref{ck}).

\begin{table}[t]
{\footnotesize
 \noindent
\caption{\small{  Left- and right-constants of the motion for  a prolonged  Poisson--Hopf  deformation of a Lie--Hamilton system   coming from a   Casimir ${C}_z$. By construction, there is  a maximal number of   $(2m-1)$ functionally independent   constants of the motion since $ F_{z{(m+1)}}\equiv  F_z^{(m+1)}$.}}
\label{table1}
\medskip
\noindent\hfill
$$
\begin{array}{cl}
\hline
&\\[-6pt]
\multicolumn{1}{c}\rm{Set of $m$ left-constants $F_z^{(k)}$ in involution}\quad&  \mbox{Tensor product space for the coproduct}\\[4pt]
F_z^{(2)} :=D_z^{(2)}\bigl[ \Delta_z^{(2)}({  C_z}) \bigr]
&   1\otimes 2 \\[4pt]  
 F_z^{(3)} :=D_z^{(3)}\bigl[ \Delta_z^{(3)}({  C_z}) \bigr] &   1\otimes 2\otimes 3\cr 
   \vdots    &\qquad \vdots \\[2pt]  
    F_z^{(k)} :=D_z^{(k)}\bigl[ \Delta_z^{(k)}({  C_z}) \bigr]  & 1\otimes 2\otimes
\ldots
\otimes k\cr
   \vdots    &\qquad \vdots \\[2pt]  
 F_z^{(m+1)} :=D_z^{(m+1)}\bigl[ \Delta_z^{(m+1)}({  C_z}) \bigr]  &  1\otimes 2\otimes
\ldots \otimes m
\otimes (m+1) \\[8pt]
\multicolumn{1}{c}\rm{Set of $m$ right-constants $F_{z{(k)}}$ in involution}\quad&  \mbox{Tensor product space  for the coproduct}\\[4pt]
F_{z{(2)}} :=D_{zR}^{(2)}\bigl[ \Delta_{zR}^{(2)}({  C_z}) \bigr]  & m\otimes (m+1) \\[4pt]   
F_{z{(3)}} :=D_{zR}^{(3)}\bigl[ \Delta_{zR}^{(3)}({  C_z}) \bigr]  & (m-1)\otimes m
\otimes (m+1)\cr
  \vdots    &\qquad \vdots \\[2pt]  
F_{z{(k)}} :=D_{zR}^{(k)}\bigl[ \Delta_{zR}^{(k)}({  C_z}) \bigr] &
(m-k+2)\otimes (m-k+3)\otimes \ldots\otimes (m+1) \cr
   \vdots    &\qquad \vdots \cr  
F_{z{(m+1)}} =  F_z^{(m+1)}:=D_{zR}^{(m+1)}\bigl[ \Delta_{zR}^{(m+1)}({  C_z}) \bigr]  &
1\otimes 2\otimes 
\ldots
\otimes m\otimes (m+1)\\[6pt]
 \hline
 \end{array}
$$
\hfill}
\end{table}



\section{Oscillator Lie--Hamilton systems}
\label{s3}

LH systems on the manifold $\mm\equiv \mathbb R^2$  were fully classified  in~\cite{BB2015}, basing on a previous classification of Lie algebras of vector fields in the real plane obtained in~\cite{GKO}. It turns out that there are 12 equivalence classes of finite-dimensional Lie algebras of Hamiltonian vector fields on  $\mathbb R^2$. For most of these planar LH systems, the constants of the motion and the superposition rules were inspected in~\cite{Ba2015}. The simple Lie algebra $\mathfrak{sl}(2,\mathbb R)$,  that appears three times in the classification, has been studied in detail from both the non-deformed and deformed viewpoints (see e.g.~\cite{BCFHL,BCFHLb,BB2015,Ba2015}). In this section, we focus on the physically relevant    oscillator $\mathfrak{h}_4$-LH systems on $\mathbb R^2$, reviewing the main results and applications,  with the aim of introducing its Hopf algebra deformation in Section \ref{s4}, where both deformed constants of the motion and  deformed superposition rules will be determined, as an illustration of the refinement of the deformation procedure presented above.

Let us consider the class ${\rm I}_8$ in the classification of real  Lie algebras of Hamiltonian vector fields with global coordinates $\>x=\{x_1,x_2\}\equiv \{x,y\}$  on $\mathbb{R}^2$ obtained in~\cite{BB2015}. The  Vessiot--Guldberg Lie algebra  $V$ is spanned by three generators
\be
\>X_1=\frac{\partial}{\partial x},\qquad \>X_2=\frac{\partial}{\partial y},\qquad \>X_3=x\, \frac{\partial}{\partial x}- y\, \frac{\partial}{\partial y}, 
  \label{ea}
\ee
satisfying the Lie brackets
\be
[\>X_1,\>X_2]=0,\qquad [\>X_1,\>X_3]=\>X_1,\qquad  [\>X_2,\>X_3]=-\>X_2 .
  \label{eb}
\ee
Hence $V$ is isomorphic to the $(1+1)$-dimensional Poincar\'e algebra ${\mathfrak{iso}}(1,1)$.  The Lie system $\>X$   (\ref{aabb}) is given by
\be
{\bf X}(t,x,y)=  b_1(t) \, \frac{\partial}{\partial x}+  b_2(t) \, \frac{\partial}{\partial y}+ b_3(t) \, \left( x\, \frac{\partial}{\partial x}- y\, \frac{\partial}{\partial y}\right) ,
\label{ec} 
\ee
 leading to the  following first-order system 
\bea
&&\frac{\dd x}{\dd t}= b_1(t) + b_3(t) x ,\nonumber\\[2pt]
&&\frac{\dd y}{\dd t}= b_2(t) - b_3(t)  y .
\label{ed}
\eea
The generators $\>X_i$ defined in (\ref{ea}) become Hamiltonian vector fields $h_i$  with respect to the standard symplectic form
\be
\omega = \dd x \wedge \dd y,
\label{ee2}
\ee
which, after application of (\ref{ae}), are found to be
\be
h_1=y,\qquad h_2=-x,\qquad h_3= x y,\qquad h_0 =1.
\label{ee}
\ee
Note that the addition of a central generator $h_0$ is required to ensure that the corresponding brackets close as a Lie algebra:
\be
\{h_1,h_2\}_\omega=h_0,\qquad \{h_1,h_3\}_\omega=-h_1,\qquad \{h_2,h_3\}_\omega=h_2,\qquad  \{h_0,\cdot\}_\omega=0 .
\label{ef}
\ee
It follows that the resulting LH algebra ${\cal H}_\omega$  is isomorphic to the centrally extended Poincar\'e algebra $\overline{\mathfrak{iso}}(1,1)$,  which is also isomorphic  to the   oscillator  algebra $\mathfrak{h}_4$.    In particular,   we consider the usual basis of $\mathfrak{h}_4=\{A_-,A_+,N,I\}$   corresponding to the ladder, number and central generators, respectively.   Under the identification
\be
A_-=h_1,\qquad A_+=h_2,\qquad N=-h_3,\qquad I=h_0 ,
\label{eg}
\ee
it is easily verified that the relations (\ref{ef}) are brought into the usual form for $\mathfrak{h}_4$:
\be
\{N,A_\pm\}_\omega=\pm A_\pm,\qquad \{A_-,A_+\}_\omega=I,\qquad  \{I,\cdot\}_\omega=0 .
\label{eh}
\ee
In the following we shall denote the oscillator LH algebra ${\cal H}_\omega$ (\ref{ef})  by $\h4$.


\subsection{Constants of the motion and superposition rules}
\label{s31}

We now proceed to compute   $t$-independent constants of the motion for the $\mathfrak{h}_4$-LH systems and deduce the corresponding superposition rules.

The starting point is to consider the PH algebra $C^\infty(\mathcal{H}^*_\omega)\equiv C^\infty( \h4^\ast)$   in a basis $\{ v_1,v_2,v_3,v_0\}$  satisfying the same Poisson brackets ({\ref{ef}). Now, besides $v_0$,  there exists a  non-trivial Casimir element  given by 
\be
C=v_{1}v_{2}+ v_{3}v_{0}.
\label{fa}
\ee
From $C$, applying the morphism $D: C^\infty(\h4^\ast)\to C^\infty( \mathbb R^2)$ (\ref{ch}) to the function $F$ in (\ref{cj}), where $h_i$ are given   in (\ref{ee}), we find that the constant of the motion is trivial:
\be
  F=D(C)=  h_{1}(x_1,y_1)h_{2}(x_1,y_1) +  h_{3}(x_1,y_1)h_{0}(x_1,y_1)   = -y_1x_1+ x_1 y_1\times 1=0 .
  \label{fb}
 \ee
As the index $m+1$  in (\ref{cj}) equals  $3$ (see~\cite{Ba2015}), we have that $k=2,3$. By making use of the morphisms $D^{(k)}$ in (\ref{ch}) and the coproducts $\Delta ^{(k)}$ in (\ref{cd}), we recursively construct the constants of the motion $F^{(2)}$ and $F^{(3)}$ with the aid of the functions  $h^{(k)}_{i}$ (\ref{ch}):
\bea
F^{(2)}=D^{(2)}\bigl[ \Delta ^{(2)}(C)\bigr]  \! \!\!\!& \! =\! &\! \!\!   \left( h_{1}(x_1,y_1)+h_{1}(x_2,y_2)\right) \left( h_{2}(x_1,y_1)+h_{2}(x_2,y_2)\right)\nonumber\\[2pt]
&&  + \left( h_{3}(x_1,y_1)+h_{3}(x_2,y_2)\right)
 \left( h_{0}(x_1,y_1)+h_{0}(x_2,y_2)\right)\nonumber\\[2pt]
\!\!\!& \! =\! &\! \!\!  - (  y_1+ y_2)   ( x_1+ x_2 )+(x_1 y_1 + x_2 y_2)(1+1)\nonumber\\[2pt]
\!\!\!& \! =\! &\! \!\!  (x_{1}-x_{2})   (y_{1}-y_{2}) .
 \label{fc}
 \eea
In the same way,   $F^{(3)}$ is found to be
 \bea
 &&F^{(3)}=D^{(3)}\bigl[ \Delta ^{(3)}(C)\bigr]=h^{(3)}_{1} h^{(3)}_{2} +h^{(3)}_{3}  h^{(3)}_{0}\nonumber\\[2pt]
   &&\qquad\qquad\qquad\qquad\quad\ \
 =(x_{1}-x_{2})   (y_{1}-y_{2})+(x_{1}-x_{3})   (y_{1}-y_{3})+(x_{2}-x_{3})   (y_{2}-y_{3})\nonumber\\[2pt]
 &&\qquad\qquad\qquad\qquad\quad\ \ =(2 x_1-x_2-x_3)y_1+(2 x_2-x_1-x_3)y_2+(2 x_3-x_1-x_2)y_3.
 \label{fd}
 \eea
 The elements  $F^{(2)}$ and  $F^{(3)}$ are left-constants of the motion for the diagonal prolongation $\widetilde{ \>X}^3$   to $(\mathbb{R}^2)^3$
  of  the LH system ${ \>X}$ (\ref{ec}). It can be checked that they are in involution in  $C^\infty\bigr( (\mathbb{R}^2)^3 \bigl)$, that is, they Poisson-commute with respect to the symplectic form
   \be
\omega = \dd x_1\wedge \dd y_1+\dd x_2\wedge \dd y_2+\dd x_3\wedge \dd y_3 .
\label{fe}
\ee
From $F^{(2)}$ we   obtain two additional constants of the motion through the permutations (\ref{ck}) (recall that $k=2,3$):
 \be 
    F_{13}^{(2)}=S_{13} \bigl( F^{(2)}   \bigr)=(x_{3}-x_{2})   (y_{3}-y_{2}) ,\qquad 
      F_{23}^{(2)}=S_{23} \bigl( F^{(2)}   \bigr)=(x_{1}-x_{3})   (y_{1}-y_{3}) .
\label{ff}
\ee
The remaining transposition is discarded, as $F_{12}^{(2)}=S_{12} \bigl( F^{(2)}   \bigr)\equiv F^{(2)}$. As far as the  right-constants of the motion $F_{(k)}$ are concerned, we have that $F_{(2)}\equiv  F_{13}^{(2)}$ and $F_{(3)}\equiv   F^{(3)}$ which are also in involution, while $ F_{23}^{(2)}$ does not belong to any of the sets $\left\{F^{(k)}\right\}$, $\left\{F_{(k)}\right\}$.
 
The  functions  (\ref{fc}), (\ref{fd}) and (\ref{ff}) determine four $t$-independent constants of the motion for $\widetilde{\> X}^3$. They can also be considered as $t$-independent constants of the motion for $\> X$ (\ref{ec}). Moreover, there exist some constants $k_i$ such that 
\be
F^{(2)}=  k_1,\qquad  F_{23}^{(2)}=k_2,\qquad  F_{13}^{(2)}=k_3 ,\qquad F^{(3)}=F^{(2)}+F_{23}^{(2)}+F_{13}^{(2)}=  k_1+k_2+k_3\equiv k.
\label{fg}
\ee
With these results, we can explicitly derive a superposition rule. We recall that in~\cite{Ba2015}   this was carried out by choosing $F^{(2)}$ and $F_{23}^{(2)}$, thus   expressing  $(x_1,y_1)$ in terms of $(x_2,y_2,x_3,y_3)$ and  $k_1,k_2$.  The resulting expression was further simplified by also introducing $ k_3$, explicitly
\bea
&&x_1(x_2,y_2,x_3,y_3,k_1,k_2,k_3)  = \frac 12(x_2+x_3) +\frac{k_2-k_1\pm B}{2(y_2-y_3)}   ,\nonumber\\[2pt]
&&y_1(x_2,y_2,x_3,y_3,k_1,k_2,k_3)  =\frac 12(y_2+y_3) +\frac{k_2-k_1\mp B}{2(x_2-x_3)} ,\label{fh}\\[2pt] 
&& B= \sqrt{ k_1^2+k_2^2+k_3^2-2(k_1k_2+k_1k_3+k_2k_3) } \, ,\nonumber
\eea
where $k_3=k_3(x_2,y_2,x_3,y_3)$ through $F_{13}^{(2)}$ in (\ref{ff}), and such that the following inequality is satisfied:
\be
k_1^2+k_2^2+k_3^2\geq 2(k_1k_2+k_1k_3+k_2k_3).
\label{fi}
\ee

As we shall prove in Subsection~\ref{s41},  the constant $F_{23}^{(2)}$ will disappear under the deformation, implying that we only have to consider  the three remaining (left- and right-) constants of the motion $F^{(2)}$, $F^{(3)}\equiv F_{(3)}$ and $F_{(2)}\equiv F_{13}^{(2)}$ of  (\ref{fg}). Furthermore, it will turn out that the latter relation between the right-constant of the motion and the permuted one  does not hold any more in this form. Therefore, in order to obtain a superposition rule that is consistent with the limit $z\to 0$ of the corresponding rule for the prolonged deformation of our LH system, we have to proceed without using $F_{23}^{(2)}$. To this extent, we start with $F^{(2)}$ and $F^{(3)}$, now writing $(x_1,y_1)$ in terms of $(x_2,y_2,x_3,y_3)$ and the constants $k_1$ and $k$ (instead of $k_2$). Next we introduce the constant $k_3$ to simplify the superposition rule, so that we are led to the expressions 
\bea
&&x_1(x_2,y_2,x_3,y_3,k_1,k,k_3)  = x_3  +\frac{k-2k_1\pm B}{2(y_2-y_3)}   ,\nonumber\\[2pt]
&&y_1(x_2,y_2,x_3,y_3,k_1,k,k_3)  =y_3 +\frac{k-2k_1\mp B}{2(x_2-x_3)} ,\label{fj}\\[2pt] 
&& B= \sqrt{ \bigl( k- 2 (k_1+k_3)\bigr)^2  - 4 k_1 k_3 } \, ,\nonumber
\eea
where again $k_3=k_3(x_2,y_2,x_3,y_3)$ through $F_{(2)}\equiv F_{13}^{(2)}$, subjected to the constraint
\be
\bigl( k- 2 (k_1+k_3)\bigr)^2 \geq 4 k_1 k_3.
\label{fk}
\ee
We remark that by introducing $k=k_1+k_2+k_3$  and  $k_3=(x_{3}-x_{2})   (y_{3}-y_{2})$ in (\ref{fj}), we easily recover the formulae (\ref{fh}).


\subsection{The book algebra and Lie--Hamilton systems}
\label{s32}

We observe from  (\ref{eh})  that the generator  $N$, along with either $A_+$ or $A_{-}$, span  a two-dimensional subalgebra of $\mathfrak{h}_4$  isomorphic to the  so-called   `book' algebra $\mathfrak{b}_2$, where $N$ can be see as a  dilation  and    $A_\pm$ as a translation.  In the basis of the LH algebra  $\mathfrak{h}_4$ with commutators (\ref{eb}),   we choose  the subalgebra $\mathfrak{b}_2$  as the one  generated by $\>X_2$ and $\>X_3$   :
      \be
      \>X_2=\frac{\partial}{\partial y},\qquad \>X_3=x\, \frac{\partial}{\partial x}- y\, \frac{\partial}{\partial y}, \qquad  [\>X_2,\>X_3]=-\>X_2 .
\label{ga}    
\ee
When $\mathfrak{b}_2$ is seen as a Vessiot--Guldberg Lie algebra, it gives rise to the particular Lie subsystem of (\ref{ec}) with $b_1(t)\equiv 0$:    
\bea
&&\frac{\dd x}{\dd t}=  b_3(t) x ,\nonumber\\[2pt]
&&\frac{\dd y}{\dd t}= b_2(t) - b_3(t)  y .
\label{gb}
\eea
The symplectic form (\ref{ee2}) is kept invariant, while the Hamiltonian vector fields (\ref{ee}) for $\mathfrak{b}_2$ are given by 
\be
h_2=-x,\qquad h_3= xy,\qquad \{h_2,h_3\}_\omega=h_2 .
\label{gc}
\ee
We recall   that $\mathfrak{b}_2$ arises within the classification of planar   LH systems~\cite{BB2015,Ba2015} as the class I$_{14A}^{r=1}\simeq \mathbb{R} \ltimes \mathbb{R}\simeq \mathfrak{b}_2$. Although $\mathfrak{b}_2$ does not admit  non-constant Casimir invariants,  its consideration as a particular case of the $\mathfrak{h}_4$-LH systems allows us to apply the above results concerning constants of the motion and superposition rules, as it was pointed out in~\cite{Ba2015}. Furthermore, in spite of the apparently naive form of the differential equations (\ref{gb}), it is worthy to be remarked that $\mathfrak{b}_2$-LH systems emerge in various physical and  mathematical contexts such as~\cite{BB2015,Ba2015}:
\begin{itemize}
\item {\em Generalised Buchdahl equations}, which are second-order differential equations  appearing  in the study of relativistic fluids~\cite{Buchdahl,Chandrasekar} and  have also been  studied by means of a Lagrangian approach in~\cite{Nikiciuk}.

\item Some particular two-dimensional {\em  Lotka--Volterra systems   with $t$-dependent coefficients}~\cite{Tsvetkov,Jin}.

\item {\em Complex Bernoulli differential equations with $t$-dependent real coefficients}~\cite{Muriel}, which  are the particular case of the  
non-autonomous complex Bernoulli differential equations with  complex coefficients~\cite{Zoladek, Marino}.
  
\end{itemize}

In what follows, we focus on the third type of  $\mathfrak{b}_2$-systems and its PH deformation will be obtained  in     Subsection~\ref{s43}. The two remaining types can also be developed in similar manner, although computations are rather cumbersome  due to the complicated symplectic structure that arises, as well as the  change of variables required to relate such systems to the expressions  (\ref{ga})--(\ref{gc}).


\subsection{Complex Bernoulli differential equations}
\label{s33}

Let us consider  the family of non-autonomous complex Bernoulli differential equations  
\be
 \frac{\dd w}{\dd t}=a_1(t)w+a_2(t)w^s,\qquad s\notin\{0,1\},
 \label{ha}
\ee
where $w$ is a complex function and $a_1(t),a_2(t)$ are arbitrary {\em real} valued $t$-dependent functions. Introducing the polar reference $w= r{\rm e}^{{\rm i}\theta}$, we obtain that the differential equation (\ref{ha})  unfolds as the real first-order system 
     \bea
&&\frac{\dd r}{\dd t}=  a_1(t) r+a_2(t) r^s\cos[\theta(s-1) ],\nonumber\\[2pt]
&&\frac{\dd \theta}{\dd t}= a_2(t)  r^{s-1} \sin[\theta(s-1) ],
\label{hb}
\eea
which can be expressed through the $t$-dependent vector field
\be
     {\bf Y}(t,r,\theta)= a_1(t) {\mathbf Y}_1+a_2(t) {\mathbf Y}_2,
\label{hc}
\ee
where
\be
{\>Y}_1=r\frac{\partial}{\partial r},\qquad  {\>Y}_2=r^s\cos[\theta(s-1)]\frac{\partial}{\partial r}+r^{s-1}\sin[\theta(s-1)]\frac{\partial}{\partial \theta} .
\label{hd}
\ee
The corresponding Lie bracket
\be
[{\>Y}_1, {\>Y}_2]=(s-1) {\>Y}_2,
\label{he}
\ee
shows that $  {\bf Y}$   is a Lie system with Vessiot--Guldberg Lie algebra $V$ isomorphic to $\mathfrak{b}_2$. 

The next step is to determine a  symplectic form $\omega=f(r,\theta)\dd r\wedge \dd \theta$ compatible with the vector fields   (\ref{hd})   by requiring  the relation  (\ref{ad}) to be satisfied. A routine computation shows that  $\omega$ can be chosen as
 \be
 \omega = \frac{s-1}{r  \sin^2[\theta(s-1) ]}\,  \dd r\wedge \dd \theta .
 \label{hf}
 \ee
 Therefore  $  {\bf Y}$  (\ref{hc})  is a LH system whose  Hamiltonian functions $\bar h_i$, deduced by means of the relation (\ref{ae}), are given by 
\be
\bar h_1= -\frac{1}{\tan[\theta(s-1)]} \, , \qquad \bar h_2= -\frac{r^{s-1}}{\sin[\theta(s-1)]}\,  .
 \label{hg}  
\ee
The Poisson bracket with respect to the symplectic form (\ref{hf}) reads 
\be
\{ \bar h_1,\bar h_2\}_\omega=-(s-1)\bar h_2 .
\label{hi}
\ee
     
Now our task consists in establishing the relationship of  these results  with   (\ref{ga})--(\ref{gc}). This is done considering the change of variables given
by
\bea
&& x=\frac{r^{s-1}}{\sin[\theta(s-1)]} \,  ,\qquad y=-\frac{\cos[\theta(s-1)]}{(s-1)r^{s-1}}  \, ,\nonumber\\[2pt]
&&r^{2(s-1)}=\frac{x^2}{1+(s-1)^2 x^2 y^2}  \,  ,\qquad \tan^2[\theta(s-1)]=\frac{1}{(s-1)^2 x^2 y^2}   \,  .
\label{hj}
\eea
Under these transformations, the symplectic form (\ref{hf}) adopts the canonical form (\ref{ee2}), while the relations amongst vector fields and $t$-dependent coefficients are given by
\bea
&&\>Y_1=(s-1)\>X_3,\qquad \>Y_2=\>X_2,\qquad \bar h_1=(s-1) h_3,\qquad \bar h_2=h_2,  \nonumber\\[2pt]
&&  a_1(t)=b_3(t)/(s-1) ,\qquad a_2(t)=b_2(t) .
\label{hk}
\eea 
With the relations (\ref{hj}) at hand, it is straightforward to obtain the constants of the motion for the complex Bernoulli differential equations  (\ref{hb}). The three functions $F^{(2)}$, $F^{(3)}$ and $F_{(2)}\equiv  F_{13}^{(2)}$ (see (\ref{fc}), (\ref{fd}) and (\ref{ff}) respectively)   have the explicit form 
\bea
&&\!\!\!\!\!\! \!\!\!\!\!\!
F^{(2)}=\frac{1}{1-s}\left(    \frac{r_1^{s-1}}{\sin[\theta_1(s-1)]} - \frac{r_2^{s-1}}{\sin[\theta_2(s-1)]}    \right)   \left(  \frac{\cos[\theta_1(s-1)]}{r_1^{s-1}}-\frac{\cos[\theta_2(s-1)]}{ r_2^{s-1}}    \right),\nonumber\\[2pt]
&&\!\!\!\!\!\! \!\!\!\!\!\!
 F_{(2)}=\frac{1}{1-s}\left(    \frac{r_2^{s-1}}{\sin[\theta_2(s-1)]} - \frac{r_3^{s-1}}{\sin[\theta_3(s-1)]}    \right)   \left(  \frac{\cos[\theta_2(s-1)]}{ r_2^{s-1}}-\frac{\cos[\theta_3(s-1)]}{ r_3^{s-1}}    \right),\label{hl}\\[2pt]
&&\!\!\!\!\!\! \!\!\!\!\!\!
 F^{(3)}=\frac{1}{1-s}\sum_{1\le i<j}^3\left(    \frac{r_i^{s-1}}{\sin[\theta_i(s-1)]} - \frac{r_j^{s-1}}{\sin[\theta_j(s-1)]}    \right)   \left(  \frac{\cos[\theta_i(s-1)]}{ r_i^{s-1}}-\frac{\cos[\theta_j(s-1)]}{ r_j^{s-1}}    \right) .\nonumber
\eea
These are the (left- and right-) constants  of the motion  used    to derive the superposition rule for  $\mathfrak{h}_4$-LH systems (\ref{fj}); for the 
   Bernoulli equations    this corresponds to express $(r_1,\theta_1)$  in terms of $(r_2,\theta_2,r_3,\theta_3)$  and the constants $k_1,k$. By introducing (\ref{hj}) in (\ref{fj}), we directly infer the superposition rule for the complex Bernoulli differential equations (\ref{hb}) in implicit form:
 \bea
&&\frac{r_1^{s-1}}{\sin[\theta_1(s-1)]} = \frac{r_3^{s-1}}{\sin[\theta_3(s-1)]}  +(1-s)\, \frac{k-2k_1\pm B}{2\left(\frac{\cos[\theta_2(s-1)]}{ r_2^{s-1}} -\frac{\cos[\theta_3(s-1)]}{ r_3^{s-1}} \right)}  \,  ,\nonumber\\[2pt]
&&\frac{\cos[\theta_1(s-1)]}{ r_1^{s-1}}  =\frac{\cos[\theta_3(s-1)]}{ r_3^{s-1}}+(1-s)\, \frac{k-2k_1\mp B}{2\left(\frac{r_2^{s-1}}{\sin[\theta_2(s-1)]} -\frac{r_3^{s-1}}{\sin[\theta_3(s-1)]} \right)} \, ,\label{hm} 
\eea
where the  constants  $k_1,k,k_3$ and the function $B$ are the same as in (\ref{fj}).


\section{Prolonged deformations of    oscillator Lie--Hamilton  systems}
\label{s4}

Multiparametric coboundary Lie bialgebras for the oscillator Lie algebra $\mathfrak{h}_4=\{A_-,A_+,N,I\}$    (\ref{eh}) were classified in~\cite{h4} along with their quantum deformations. This exhaustive analysis shows that mathematical and physical properties of each deformation are in direct correspondence with the generators that remain undeformed, that is, with a primitive (trivial)  coproduct (\ref{cb}). As the central generator $I$ is always primitive, one should additionally require either $N$ or a single $A_\pm$ to be primitive as well. It turns out that all  (multiparametric)  deformations with $N$ primitive lead to quantum deformations that are governed by $I$~\cite{h4},  with $N$  behaving as a `secondary' primitive generator. In the context  of LH systems this implies that  these  quantum deformations give rise to `trivial'  LH systems  (recall that $I=h_0=1$ in (\ref{eg})). By contrast, deformations with a   primitive $A_+=-x$ (or $A_-=y$) provide non-trivial LH systems, as in these cases $A_+$ plays the role of the `main' primitive generator, with $I$ playing the role of a  `secondary' one.
   
The simplest (i.e.\ one-parameter) quantum deformation such that $A_+$ is primitive corresponds to consider the classical $r$-matrix
\be
r=  z\, A_+\wedge   N,
\label{ia}
\ee
 which is a solution of the classical Yang--Baxter equation, and  where $ z$ is the quantum deformation parameter such that $q={\rm e}^z$. 
 This element underlies the so-called nonstandard (or Jordanian) quantum oscillator algebra $U_{  z}(\mathfrak{h}_4)$, whose boson representations have been studied in~\cite{boson,boson2}.

In the LH framework, we start with the Lie algebra $\mathfrak{h}_4$  in the basis $\{ v_1,v_2,v_3,v_0\}$ with Lie brackets (see (\ref{ef}))
\be
[v_1,v_2] =v_0,\qquad [v_1,v_3]=-v_1,\qquad [v_2,v_3] =v_2,\qquad  [v_0,\cdot]=0 ,
\label{ib}
\ee
as well as with the classical $r$-matrix
\be
r=z\, v_3\wedge v_2 .
\label{ic}
\ee
  The Lie bialgebra is provided by the cocommutator map $\delta$ that is obtained from the classical $r$-matrix as
  \begin{equation}
\delta(v_i)=[v_i\otimes 1+1\otimes v_i , r], 
\label{idd}
\end{equation}
yielding
\be
\delta(v_2)=\delta(v_0)=0,\qquad \delta(v_1)=z(v_2\wedge v_1+ v_3\wedge v_0),\qquad  \delta(v_3)=z\, v_2\wedge v_3,
\label{id}
\ee
which is just the  skew-symmetric part of the first-order term $\Delta_1$ in $z$  of the full coproduct $\Delta_z$, that is,
\be
\Delta_z(v_i)=\Delta_0(v_i)+ \Delta_1(v_i)+o[z^2], \quad\   \Delta_0(v_i)=v_i\otimes 1+ 1\otimes v_i ,\quad\  \delta(v_i)=\Delta_1(v_i)-\sigma\circ \Delta_1(v_i) ,
\label{ie}
\ee
where $\sigma$ is the flip operator: $\sigma(v_i\otimes v_j)=v_j\otimes v_i$.

From the complete quantum algebra $U_{   z}(\mathfrak{h}_4)$~\cite{h4,boson},  the  corresponding Poisson coalgebra structure can easily be  deduced giving rise to the following deformed coproduct and Poisson brackets:
\bea
&& \Delta_z(v_2)=v_2\otimes 1 + 1 \otimes v_2,\qquad  \Delta_z(v_0)=v_0\otimes 1 + 1 \otimes v_0,\nonumber\\[2pt]
&&
\Delta_z(v_1)=v_1\otimes \eee^{-z v_2}+1\otimes v_1 + z \, v_3 \otimes \eee^{-z v_2} \, v_0 ,\qquad\Delta_z(v_3)=v_3\otimes \eee^{-z v_2}+1\otimes v_3 ,
\label{if}
\eea
\be
\{v_1,v_2\}_z =\eee^{-z v_2}\,v_0,\qquad \{v_1,v_3\}_z=-v_1,\qquad \{v_2,v_3\}_z=\frac{1-\eee^{-z v_2}}{z},\qquad  \{v_0,\cdot\}_z=0 ,
\label{ig}
\ee
such that $\Delta_z$   (\ref{if}) satisfies the coassociativity condition  (\ref{ba}) and   is a Poisson algebra homomorphism of the Poisson brackets (\ref{ig}).
The deformed Casimir   turns out to be
\be
C_z=v_1\left(  \frac{\eee^{z v_2}-1}{z}  \right) +v_3v_0 .
\label{ih}
\ee

Now we apply the algorithmic procedure summarized in Subsection~\ref{s22} to construct a PH deformation $C^\infty(\hz^\ast)$ of $C^\infty(\h4^\ast)$, hence  deforming   the $\mathfrak{h}_4$-LH    systems of Section~\ref{s3}. To this extent, we start from the functions $\{h_1,h_2,h_3,h_0\}$ on $C^\infty( \mathbb R^2)$ as given in (\ref{ee}) with Poisson brackets (\ref{ef}),  where $\omega$ is the canonical symplectic form (\ref{ee2}). Taking into account the  boson representations of $U_{  z}(\mathfrak{h}_4)$ given in~\cite{boson,boson2},  
we introduce the Hamiltonian functions on $C^\infty(\mathbb R^2)$
\be
h_{z,1}= \eee^{z x} y,\qquad h_{z,2}= -x,\qquad  h_{z,3}=\left( \frac{ \eee^{z x} -1}{z}\right) y,\qquad   h_{z,0}=1,
\label{ii}
\ee
which  satisfy the following Poisson brackets with respect to the same symplectic form (\ref{ee2})
\bea
&&\{h_{z,1} ,h_{z,2}\}_\omega =\eee^{-z h_{z,2}}\, h_{z,0},\qquad \{h_{z,1} ,h_{z,3}\}_\omega=-h_{z,1},\nonumber\\[2pt]
&& \{h_{z,2},h_{z,3}\}_\omega=\frac{1-\eee^{-z h_{z,2}}}{z},\qquad  \{ h_{z,0},\cdot\}_\omega=0 ,
\label{ij}
\eea
  in agreement with the relations (\ref{ig}).  In the third step, the deformed vector fields $\>X_{z,i}$ on $\mathbb R^2$ are obtained through the relation (\ref{bg}), namely
 \be
\>X_{z,1}=\eee^{z x} \frac{\partial}{\partial x}  -z  \eee^{z x} y \,  \frac{\partial}{\partial y}  ,\qquad \>X_{z,2}= \frac{\partial}{\partial y} ,\qquad  \>X_{z,3}=\left( \frac{ \eee^{z x}-1}{z} \right)\frac{\partial}{\partial x}- \eee^{z x} y\, \frac{\partial}{\partial y} \, .
\label{ik}
\ee
Finally, the PH deformation of the $\mathfrak{h}_4$-LH    system  (\ref{ec}) is determined by 
\be
{\bf X}_z(t,x,y)=  b_1(t) \,\>X_{z,1}+  b_2(t) \, \>X_{z,2}+ b_3(t) \,\>X_{z,3},
\label{il} 
\ee
leading to the system of  differential equations  
\bea
&&\frac{\dd x}{\dd t}= b_1(t) \,\eee^{z x}+ b_3(t) \left( \frac{ \eee^{z x}-1}{z} \right) ,\nonumber\\[2pt]
&&\frac{\dd y}{\dd t}= b_2(t) - \bigr( b_3(t)   + z\, b_1(t)  \bigl)    \eee^{z x} y .
\label{im}
\eea
It is worth remarking that since  $\omega$ is the standard symplectic form (\ref{ee2}),  the same system of  differential equations   (\ref{im}) can, alternatively,  be obtained by computing the usual  Hamilton equations from the deformed Hamiltonian
(\ref{bk}) with the functions (\ref{ii}),  
 \be
h_{z} =b_1(t)  \eee^{z x}y - b_2(t)   x + b_3(t)  \left( \frac{ \eee^{z x} -1}{z}\right) y + b_0(t),
\label{im2}
\ee
in the form
\be
\frac{\dd x}{\dd t}=\frac{\partial h_{z}}{\partial y},\qquad \frac{\dd y}{\dd t}=-\frac{\partial h_{z}}{\partial x}.
\label{im32}
\ee

  As we have already commented, 
the deformed vector fields (\ref{ik}) span a Stefan--Sussman distribution~\cite{Vaisman,Pa57,WA} whose commutation rules (\ref{bj}) turn out to be
\be
[{\bf X}_{z,1},{\bf X}_{z,2}]= z\, \eee^{-z h_{z,2}} \,h_{z,0} \,{\bf X}_{z,2},\qquad [{\bf X}_{z,1},{\bf X}_{z,3}]=   {\bf X}_{z,1}, \qquad [{\bf X}_{z,2},{\bf X}_{z,3}]= - \eee^{-z h_{z,2}} \, {\bf X}_{z,2}.
\label{in}
\ee
By introducing the functions (\ref{ii}) we   obtain  that
\be
[{\bf X}_{z,1},{\bf X}_{z,2}]= z\, \eee^{z x} \, {\bf X}_{z,2},\qquad [{\bf X}_{z,1},{\bf X}_{z,3}]=   {\bf X}_{z,1}, \qquad [{\bf X}_{z,2},{\bf X}_{z,3}]= - \eee^{z x} \, {\bf X}_{z,2}.
\label{inb}
\ee
Note that that   the expressions (\ref{ii})--(\ref{inb}) reduce to  (\ref{ea})--(\ref{ef}) in the limit $z\to 0$. It is worth stressing that, even considering the commutation relations (\ref{inb}) up to first-order in the deformation parameter $z$, the vector fields $\>X_{z,1},\>X_{z,2},\>X_{z,3}$ do not close on a finite-dimensional Lie algebra.

The remarkable feature of the deformation is the presence of the `interacting' term $   \eee^{z x} y$   in (\ref{im}),  when compared with (\ref{ed}). This nonlinear interaction or coupling between the two variables can be regarded as a perturbation of the initial system.  Indeed, by considering a power series expansion in $z$ of the system (\ref{im}) and truncating at the first-order, we obtain 
\bea
&&\frac{\dd x}{\dd t}= b_1(t) +\bigl( b_3(t)+z\, b_1(t) \bigr) x  +\frac 12\, z  \,   b_3(t)x^2  +o[z^2],\nonumber\\[2pt]
&&\frac{\dd y}{\dd t}= b_2(t) - \bigl( b_3(t)+z\, b_1(t) \bigr)  y -z\, b_3(t)    x y  +o[z^2] .
\label{io}
\eea
In the first equation, the deformation introduces a quadratic term $x^2$, leading to a Riccati equation with $t$-dependent real coefficients~\cite{PW}, while in the second one, we obtain the nonlinear interaction term $xy$. 


\subsection{Deformed constants of the motion and deformed superposition rules}
\label{s41}

Now we proceed to apply the approach presented  in Subsection~\ref{s232} in order  to obtain three deformed constants of the motion $F_z^{(2)}$, $F_{z{(2)}}$ and  $F_z^{(3)}\equiv F_{z{(3)}}$ (see  Table~\ref{table1})  for the prolonged deformation of   $\mathfrak{h}_4$-LH systems, as in this case (see  Subsection~\ref{s31}) we have the indices $m+1=3$ and $k=2,3$.

Before considering the coalgebra structure, we note that the morphisms $\phi_z:\hz\to C^\infty(\mathbb R^2)$ in (\ref{dc}) and $D_z: C^\infty(\hz^\ast)\to C^\infty(\mathbb R^2)$ in (\ref{dd}) lead to  
\be
D_z( v_i)= h_{z,i}(x_1,y_1):=  h_{z,i}^{(1)}  , \quad i=0,1,2,3,
\label{ja}
\ee
where $h_{z,i}$ are the Hamiltonian functions (\ref{ii}) fulfilling (\ref{ij}). By introducing this result into the Casimir (\ref{ih}), we find that, as expected, the corresponding constant  $ F_z$ of (\ref{df}) is again trivial:
\be
  F_z= D_z(C_z)=h_{z,1}^{(1)}  \left(  \frac{\eee^{z h_{z,2}^{(1)}  }-1}{z}  \right) +h_{z,3}^{(1)}   h_{z,0}^{(1)}  =0 .
  \label{jb}
  \ee
Now we consider the deformed coproduct $\Delta_z\equiv \Delta_z^{(2)}$ (\ref{if}) on the tensor product space $1\otimes 2$ and compute the elements
$ D_z^{(2)} \bigl( {\Delta}^{(2)}_z(v_i) \bigr) $ coming from the morphism $ D_z^{(2)}$ in (\ref{dd}) yielding the functions  $h_{z,i}^{(2)}$ (\ref{de2}) by means of (\ref{chxy}):
\bea 
 &&  D_z^{(2)} \bigl( {\Delta}^{(2)}_z(v_2) \bigr) = h_{z,2}(x_1,y_1)+h_{z,2}(x_2,y_2)= -x_1-x_2:= h_{z,2}^{(2)} \,  ,  \nonumber \\[2pt]
  &&  D_z^{(2)} \bigl( {\Delta}^{(2)}_z(v_0) \bigr) = h_{z,0}(x_1,y_1)+h_{z,0}(x_2,y_2)= 1+1=2 := h_{z,0}^{(2)} \,  ,  \nonumber \\[2pt]
 && 
D_z^{(2)} \bigl( {\Delta}^{(2)}_z(v_1) \bigr) = h_{z,1}(x_1,y_1)  {\rm e}^{- z h_{z,2}(x_2,y_2)}+ h_{z,1}(x_2,y_2)   + z\, h_{z,3}(x_1,y_1)   {\rm e}^{- z h_{z,2}(x_2,y_2)}h_{z,0}(x_2,y_2) \nonumber \\[2pt]
&&\qquad\qquad\qquad \ \ = \eee^{z x_1}\eee^{z x_2}  y_1 +\eee^{z x_2}  y_2 + \left(  { \eee^{z x_1} -1}\right) \eee^{z x_2}y_1   :=  h_{z,1}^{(2)} \,   ,\nonumber \\[2pt]
  && 
D_z^{(2)} \bigl( {\Delta}^{(2)}_z(v_3) \bigr) = h_{z,3}(x_1,y_1)  {\rm e}^{- z h_{z,2}(x_2,y_2)}+ h_{z,3}(x_2,y_2)    \nonumber \\[2pt]
&&\qquad\qquad\qquad \ \   =\left( \frac{ \eee^{z x_1} -1}{z}\right) \eee^{z x_2}y_1+\left( \frac{ \eee^{z x_2} -1}{z}\right) y_2
  :=  h_{z,3}^{(2)} \,  .
\label{jc}
\eea
These expressions allow us to obtain the left-constant of the motion of (\ref{df}) for $k=2$:  
\be
F_z^{(2)}=D_z^{(2)}\bigl[\Delta_z^{(2)} \bigl({C_z } \bigr) \bigr] =h_{z,1}^{(2)}  \left(  \frac{\eee^{z h_{z,2}^{(2)}  }-1}{z}  \right) +h_{z,3}^{(2)}   h_{z,0}^{(2)}    ,
\label{jd}
\ee
namely
\be
F_z^{(2)}=\left( \frac{2- \eee^{-z x_1} -\eee^{z x_2}}{z}\right)  (y_1-y_2) .
\label{je}
\ee
Similarly, the  right-constant of the motion $F_{z{(2)}}$ defined in (\ref{dk}) is deduced, but now working with the right-coproduct  $ \Delta_{zR}^{(2)}$  on the tensor product space $2\otimes 3$ such that the functions  $  h^{(2)}_{zR,i}$ (\ref{df2}) turn out to be
\bea 
 &&  D_{zR}^{(2)} \bigl( {\Delta}^{(2)}_{zR}(v_2) \bigr) = h_{z,2}(x_2,y_2)+h_{z,2}(x_3,y_3)= -x_2-x_3=: h_{zR,2}^{(2)} \,  ,  \nonumber \\[2pt]
  &&  D_{zR}^{(2)} \bigl( {\Delta}^{(2)}_{zR}(v_0) \bigr) = h_{z,0}(x_2,y_2)+h_{z,0}(x_3,y_3)= 1+1=2=: h_{zR,0}^{(2)} \,  ,  \nonumber \\[2pt]
 && 
D_{zR}^{(2)} \bigl( {\Delta}^{(2)}_{zR}(v_1) \bigr) = h_{z,1}(x_2,y_2)  {\rm e}^{- z h_{z,2}(x_3,y_3)}+ h_{z,1}(x_3,y_3)   + z\, h_{z,3}(x_2,y_2)   {\rm e}^{- z h_{z,2}(x_3,y_3)}h_{z,0}(x_3,y_3) \nonumber \\[2pt]
&&\qquad\qquad\qquad \quad\, = \eee^{z x_2}\eee^{z x_3} y_2 +\eee^{z x_3}  y_3 + \left(  { \eee^{z x_2} -1}\right) \eee^{z x_3}y_2  =:  h_{zR,1}^{(2)} \,   ,\nonumber \\[2pt]
  && 
D_{zR}^{(2)} \bigl( {\Delta}^{(2)}_{zR}(v_3) \bigr) = h_{z,3}(x_2,y_2)  {\rm e}^{- z h_{z,2}(x_3,y_3)}+ h_{z,3}(x_3,y_3)    \nonumber \\[2pt]
&&\qquad\qquad\qquad \quad\, =\left( \frac{ \eee^{z x_2} -1}{z}\right) \eee^{z x_3}y_2+\left( \frac{ \eee^{z x_3} -1}{z}\right) y_3
  =:  h_{zR,3}^{(2)} \,  ,
\label{jf}
\eea
giving rise, through the analogous expression to (\ref{jd}), to
 \be
F_{z{(2)}}=\left( \frac{2- \eee^{-z x_2} -\eee^{z x_3}}{z}\right)  (y_2-y_3) .
\label{jg}
\ee
The non-deformed limit $z\to 0$ of the expressions (\ref{je}) and (\ref{jg}) yields the functions $F^{(2)}$   (\ref{fc}) and $F_{(2)}\equiv  F_{13}^{(2)}=S_{13} ( F^{(2)} )$   (\ref{ff}).  Nevertheless,  we stress that in the deformed case, the constant of the motion $F_z^{(2)}$  does not remain invariant under the permutation $ S_{12}$ and, moreover, $F_{z{(2)}}$  is related to $F_z^{(2)}$  through the composition of two permutations which differs from the result obtained merely applying the permutation $S_{13} $:
\be
F_z^{(2)} \ne  S_{12}\bigl(F_z^{(2)}\bigr) , \qquad  F_{z{(2)}}= S_{12}\bigl( S_{23}\bigl(F_z^{(2)} \bigr) \bigr) \ne  S_{13}\bigl(F_z^{(2)}\bigr) .
\ee
In addition, there is no deformed constant of the motion that corresponds in the limit $z\to 0$ to $ F_{23}^{(2)}=S_{23} \bigl( F^{(2)} \bigr)$ in (\ref{ff}). In fact, it is straightforward to check that the functions obtained from  $F_z^{(2)}$ by means of the permutations $S_{12}$, $S_{13}$ and $S_{23}$,
\bea
&&S_{12}\bigl(F_z^{(2)}\bigr)= \left( \frac{2- \eee^{-z x_2} -\eee^{z x_1}}{z}\right)  (y_2-y_1)  ,\qquad S_{13}\bigl(F_z^{(2)}\bigr)= \left( \frac{2- \eee^{-z x_3} -\eee^{z x_2}}{z}\right)  (y_3-y_2) 
 ,\nonumber\\[2pt]
&&S_{23}\bigl(F_z^{(2)}\bigr)= \left( \frac{2- \eee^{-z x_1} -\eee^{z x_3}}{z}\right)  (y_1-y_3),
\label{jm}
\eea
do not provide any constant of the motion.  This shows that  the range of application of the permutations $S_{ij}$ in the form (\ref{ck}) is rather limited in the deformed case, where only left- and right-constants of the motion can be ensured to be correct.

It remains to compute  $F_z^{(3)}\equiv F_{z{(3)}}$, which requires to construct the third-order coproduct $\Delta_z^{(3)}$ of (\ref{if}) on the tensor product space $1\otimes 2\otimes 3$.  In this case, with $m+1=3$, $\Delta_z^{(3)}$ is obtained by means of the  coassociativity condition (\ref{ba}) (corresponding to (\ref{cd}) and (\ref{dh}) with $k=m+1=3$)
\be
\Delta_z^{(3)}=({\rm Id} \otimes \Delta_z) \circ \Delta_z =(\Delta_z \otimes {\rm Id}) \circ \Delta_z=\Delta_{zR}^{(3)} ,
\label{jh}
\ee
 leading to
 \bea
 &&
 \Delta_z^{(3)}(v_l)=v_l\otimes 1\otimes 1+ 1\otimes v_l\otimes 1+ 1\otimes 1\otimes v_l,\qquad l=0,2,\nonumber\\[2pt]
  &&  \Delta_z^{(3)}(v_1)=v_1\otimes \eee^{-z v_2}\otimes \eee^{-z v_2}+ 1\otimes v_1\otimes \eee^{-z v_2}+ 1\otimes 1\otimes v_1
 \nonumber\\[2pt]
  && \qquad\qquad\quad  +z\left( v_3\otimes \eee^{-z v_2}v_0\otimes \eee^{-z v_2}+ v_3\otimes \eee^{-z v_2}\otimes \eee^{-z v_2} v_0+   1   \otimes  v_3\otimes \eee^{-z v_2} v_0 \right), \nonumber\\[2pt]
  &&  \Delta_z^{(3)}(v_3)=v_3\otimes \eee^{-z v_2}\otimes \eee^{-z v_2}+ 1\otimes v_3\otimes \eee^{-z v_2}+ 1\otimes 1\otimes v_3,
 \eea
provided that
\be
\Delta_z(1)=1\otimes 1,\qquad \Delta_z(\eee^{-z v_2})=\eee^{-z v_2}\otimes \eee^{-z v_2} .
\label{ji}
\ee
Then, by using (\ref{chxy}),  we obtain the Hamiltonian  functions on $(\mathbb{R}^2)^3$ given by
\bea
&&\!\!\!\!\!\!\!\!
h_{z,2}^{(3)} := D_z^{(3)} \bigl( {\Delta}^{(3)}_z(v_2) \bigr) =-x_1-x_2-x_3,\qquad  h_{z,0}^{(3)} := D_z^{(3)} \bigl( {\Delta}^{(3)}_z(v_0) \bigr) =3,   \nonumber \\[2pt]
  &&\!\!\!\!\!\!\!\!	
h_{z,1}^{(3)} :=  D_z^{(3)} \bigl( {\Delta}^{(3)}_z(v_1) \bigr)  =\bigl( 3\eee^{z x_1}  -2 \bigr) \eee^{z (x_2+x_3)} y_1+ 
\bigl( 2\eee^{z x_2}  -1 \bigr) \eee^{z  x_3 } y_2+\eee^{z  x_3 } y_3 ,\nonumber \\[2pt]
  &&\!\!\!\!\!\!\!\! 
h_{z,3}^{(3)} :=  D_z^{(3)} \bigl( {\Delta}^{(3)}_z(v_3) \bigr)  =\left( \frac{ \eee^{z x_1} -1}{z}\right) \eee^{z (x_2+x_3)}y_1+\left( \frac{ \eee^{z x_2} -1}{z}\right)  \eee^{z x_3} y_2+\left( \frac{ \eee^{z x_3} -1}{z}\right) y_3  .
\label{jj}
\eea
By introducing them into  (\ref{df}) we get the third constant of the motion:
\bea
&&F_z^{(3)}=\frac 1z  \bigl(  3 -2\, \eee^{-z x_1} - \eee^{z x_2} \eee^{z x_3} \bigr) y_1+\frac 1z  \bigl(   2\, \eee^{-z x_1} -   \eee^{-z x_1} \eee^{-z x_2} 
- 2\, \eee^{z x_3} + \eee^{z x_2} \eee^{z x_3}\bigr)y_2 \nonumber\\[2pt]
&&\qquad \qquad - \frac 1z  \bigl(  3 -2\, \eee^{z x_3} - \eee^{-z x_1} \eee^{-z x_2} \bigr) y_3 ,
\label{jk}
\eea
whose limit $z\to 0$ directly gives $F^{(3)}$ in the second form written in (\ref{fd}).

Summing up, the functions (\ref{jj}) satisfy the Poisson brackets (\ref{ij})   with respect to the   symplectic form (\ref{fe}) and, with this $\omega$, all the following Poisson brackets vanish $(i=0,1,2,3)$:
\be
\bigl\{ F_z^{(2)}, h_{z,i}^{(3)}  \bigr\}_\omega=\bigl\{ F_{z{(2)}} , h_{z,i}^{(3)}  \bigr\}_\omega=\bigl\{ F_z^{(3)}, h_{z,i}^{(3)}  \bigr\}_\omega=0,\qquad 
\bigl\{ F_z^{(2)},  F_z^{(3)} \bigr\}_\omega=\bigl\{ F_{z{(2)}} ,  F_z^{(3)} \bigr\}_\omega =0 .
\label{jl}
\ee
Consequently, $F_z^{(2)}$,  $F_{z{(2)}}$ and  $F_z^{(3)}$, as given in (\ref{je}), (\ref{jg}) and (\ref{jk}), are three functionally independent constants of the motion of the prolonged   deformation $\widetilde{\>X}_z^3$ of           $\mathfrak{h}_4$-LH    systems   to $(\mathbb{R}^2)^3$. 

Let us deduce  now $\widetilde{\>X}_z^3$ in an explicit manner. By taking into account that  $\omega$  (\ref{fe}) is the standard symplectic form, we consider
  the corresponding deformed Hamiltonian on $(\mathbb{R}^2)^3$,  
\be
h_z^{(3)}=  b_1(t) h_{z,1}^{(3)} +b_2(t) h_{z,2}^{(3)} +b_3(t) h_{z,3}^{(3)}  +b_0(t) h_{z,0}^{(3)},
\label{jm2}
\ee
 with the Hamiltonian functions    (\ref{jj}), and compute the  Hamilton equations (similarly to (\ref{im32})), thus finding  that   $\widetilde{\>X}_z^3$  
 is given by the following system of six  differential equations:
  \bea
&& \frac{\dd x_1}{\dd t}=b_1(t)\bigl(3\eee^{z x_1}-2\bigr)  \eee^{z(x_2+x_3)} +b_3(t) \!\left(\frac{\eee^{zx_1}-1}{z}\right)\!\eee^{z(x_2+x_3)},    \nonumber\\[2pt]
&&\frac {\dd y_1}{\dd t}=b_2(t)-\bigl( b_3(t)+ 3z b_1(t)  \bigr) \eee^{z(x_1+x_2+x_3)} y_1,\nonumber\\[2pt]
&&\frac{\dd x_2}{\dd t}=b_1(t)\bigl(2\eee^{z x_2}-1\bigr)\eee^{zx_3}+b_3(t)\!\left(\frac{\eee^{zx_2}-1}{z}\right)\!\eee^{zx_3}, \label{jm3} \\
 && \frac {\dd y_2}{dt}=b_2(t)-b_3(t)\,\eee^{z(x_2+x_3)}\bigl((\eee^{z x_1 }-1)y_1+y_2\bigr)-zb_1(t)\,\eee^{z(x_2+x_3)}\bigl ((3\eee^{z x_1}-2)y_1+2y_2 \bigr),\nonumber\\[2pt] 
  &&\frac{\dd x_3}{\dd t}=b_1(t)\,\eee^{zx_3}+b_3(t)\left(\frac{\eee^{zx_3}-1}{z}\right),\nonumber\\[2pt]
&&  \frac {\dd y_3}{\dd t}=b_2(t) - b_3(t) \,\eee^{zx_3} \bigl(  (\eee^{zx_1}-1) \eee^{zx_2}  y_1+ (\eee^{zx_2}-1) y_2 +    y_3 \bigr)   \nonumber\\[2pt]
&&\qquad \qquad\qquad -   z b_1(t) \, \eee^{zx_3} \bigl(  (3\eee^{zx_1}-2) \eee^{zx_2}  y_1+ (2\eee^{zx_2}-1) y_2+y_3  \bigr) .\nonumber
\eea
Under the non-deformed limit $z\to 0$, the  prolonged   deformation   $\widetilde{\>X}_z^3$ reduces to  the diagonal prolongation
 $\widetilde{\>X}^3$ of   the $\mathfrak{h}_4$-LH      system $ {\>X}$ (\ref{ed})    to $(\mathbb{R}^2)^3$ which simply corresponds to three copies of  $ {\>X}$. On the contrary,   it is remarkable that   $\widetilde{\>X}_z^3$ (\ref{jm3}) is no longer formed by three copies 
of the deformed $\mathfrak{h}_4$-LH system $\>X_z$~(\ref{im}). Therefore, we stress that  the constants of the motion of $\widetilde{\>X}_z^3$ cannot be considered as constants of the motion of $\> X_z$.

 Furthermore,  the   deformed vector fields $ \> X_{h^{(3)}_{z,i}}$, which determine  $\widetilde{\>X}_z^3$ by means of the expression (\ref{de3}), 
 can directly be    deduced  from  (\ref{jm3}); these are
  \bea
 && \! \! \! \! \! \! \! \!  \> X_{h^{(3)}_{z,1}}=\bigl(3\eee^{z x_1}-2\bigr)\eee^{z(x_2+x_3)}  \frac{\partial}{\partial x_1}+ \bigl(2\eee^{z x_2}-1\bigr)\eee^{zx_3}\frac{\partial}{\partial x_2} +\eee^{zx_3}\,\frac{\partial}{\partial x_3}-  3z  \eee^{z(x_1+x_2+x_3)} y_1\,\frac{\partial}{\partial y_1}\nonumber\\ 
 &&\qquad   -z \eee^{z(x_2+x_3)}\bigl ((3\eee^{z x_1}-2)y_1+2y_2 \bigr) \frac{\partial}{\partial y_2}-   z  \eee^{zx_3} \bigl( (3\eee^{zx_1}-2)  \eee^{zx_2}  y_1+ (2\eee^{zx_2}-1) y_2+y_3  \bigr) \frac{\partial}{\partial y_3},   \nonumber\\ 
  &&\! \! \! \! \! \! \! \!  \> X_{h^{(3)}_{z,2}}= \frac{\partial}{\partial y_1}+ \frac{\partial}{\partial y_2}+ \frac{\partial}{\partial y_3},   \label{jn3}\\[4pt]
 &&\! \! \! \! \! \! \! \!  \> X_{h^{(3)}_{z,3}}= \left(\frac{\eee^{zx_1}-1}{z}\right)\!\eee^{z(x_2+x_3)} \frac{\partial}{\partial x_1}+\left(\frac{\eee^{zx_2}-1}{z}\right) \! \eee^{zx_3}\frac{\partial}{\partial x_2} +\left(\frac{\eee^{zx_3}-1}{z}\right) \!\frac{\partial}{\partial x_3}-\eee^{z(x_1+x_2+x_3)} y_1\frac{\partial}{\partial y_1}\nonumber\\ 
 &&\qquad  - \eee^{z(x_2+x_3)}\bigl((\eee^{z x_1 }-1)y_1+y_2\bigr)  \frac{\partial}{\partial y_2}-  \eee^{zx_3} \bigl(  (\eee^{zx_1}-1)  \eee^{zx_2} y_1+ (\eee^{zx_2}-1) y_2 +    y_3 \bigr)  \frac{\partial}{\partial y_3} .
 \nonumber
 \eea
 It can be checked that they      fulfil  the relationship (\ref{bg}) with respect to the Hamiltonian functions (\ref{jj}) and symplectic form (\ref{fe}).    They satisfy the commutation relations (\ref{df3}),  which are just those given in (\ref{in}) with the   functions    (\ref{jj}), that is,
 \bea
&&\left[ \> X_{h^{(3)}_{z,1}},  \> X_{h^{(3)}_{z,2}} \right]= 3z\,  \eee^{z(x_1+x_2+x_3)} \,  \> X_{h^{(3)}_{z,2}},\qquad  \left[ \> X_{h^{(3)}_{z,1}}, \> X_{h^{(3)}_{z,3}}\right]=    \> X_{h^{(3)}_{z,1}}, \nonumber\\[2pt]
&&\left[ \> X_{h^{(3)}_{z,2}}, \> X_{h^{(3)}_{z,3}}\right]= - \eee^{z(x_1+x_2+x_3)} \,  \> X_{h^{(3)}_{z,2}} ,
\label{inx}
\eea
to be compared with (\ref{inb}).

The   three deformed constants of the motion satisfying (\ref{jl}) allow us to deduce a deformed superposition rule for $\widetilde{\>X}_z^3$. 
We keep the notation of Subsection~\ref{s31} and consider the three equations coming from  (\ref{je}),    (\ref{jg})  and    (\ref{jk}):
\be
F_z^{(2)}=k_1,\qquad F_z^{(3)}=k,\qquad F_{z{(2)}}=k_3,
\label{jn}
\ee
where $k_1$, $k$ and $k_3$ are constants. From the first two equations we can express $(x_1,y_1)$ in terms of $(x_2,y_2,x_3,y_3)$ and the constants $k_1$ and $k$. The third equation
enables to write the result in a simplified manner, namely   
\bea
&&\eee^{zx_1}=\frac{   \bigl( 1+ \eee^{-z x_2}-\eee^{z x_3}  \bigr)     (y_2-y_3)+\frac z2  \bigl( 2k_1-k\pm B  \bigr)    }{z k \bigl( \eee^{z x_2}-2  \bigr) -   z k_1 \bigl( \eee^{z x_2} \eee^{z x_3}-3  \bigr) +   \bigl( \eee^{z x_2}-2  \bigr)    \bigl( 2\,\eee^{z x_3}-3  \bigr) (y_2-y_3) } ,\nonumber\\
&&y_1 =\eee^{-z x_2}y_3 +\bigl(1-\eee^{-z x_2}\bigr) y_2  +\frac{z\bigl(k-2k_1\mp B\bigr)}{2\,\eee^{z x_2}\bigl(2-\eee^{-z x_2}-\eee^{z x_3}\bigr)}  , \label{jp}\\
&&
B= \sqrt{ \bigl( k- 2 (k_1+k_3)\bigr)^2  - 4 k_1 k_3 } \, .
\nonumber
\eea
The factor $B$ is  formally the same given in (\ref{fj}), while the constant $k_3$, that only appears within $B$,  should be understood as a function   $k_3=k_3(x_2,y_2,x_3,y_3)$ through $F_{z{(2)}}$   (\ref{jg}). Therefore, the expressions (\ref{jp})  constitute a generic  deformed superposition rule corresponding to the prolonged deformation  $\widetilde{\>X}_z^3$ of      $\mathfrak{h}_4$-LH    systems   to $(\mathbb{R}^2)^3$ given by the system of  differential equations (\ref{jm3}).
 In order to compute their undeformed limit one should apply in (\ref{jp}) the limits  
\be
\lim_{z\to 0}\left( \frac{\eee^{zx_1}-1}{z} \right) ,\qquad  \lim_{z\to 0}y_1 ,
\ee
thus recovering the proper superposition rule (\ref{fj}) for  the $\mathfrak{h}_4$-LH    systems (\ref{ed}).


\subsection{Twist maps and canonical transformations}
\label{s42}

The above results illustrate how to apply the PH deformation approach from a given quantum algebra in order to obtain the corresponding deformed LH systems. The essential tool in the formalism is the coproduct map so that  one can start with either a coboundary quantum algebra (with an underlying classical $r$-matrix $r$)  or with a non-coboundary one (without  $r$-matrix). There exist two types of coboundary deformations~\cite{CP}:    quasitriangular (or standard) deformations whose $r$-matrix is a solution of the modified classical Yang--Baxter equation (like Drinfel'd--Jimbo deformations) and triangular (or nonstandard) deformations,  for which the $r$-matrix is a solution of the   classical Yang--Baxter equation (like the one  considered in this paper for $\mathfrak{h}_4$ and also in~\cite{BCFHL,BCFHLb} for $\mathfrak{sl}(2,\mathbb R)$). The latter quantum algebras  are  twist deformations~\cite{CP,Drinfelda,Drinfeld, Reshetikhin}  and  only for them it is guaranteed that  there exists a basis for which the Hopf algebra structure can be written in terms of non-deformed commutation relations and a deformed coproduct. Therefore, in our example based on the nonstandard quantum algebra $U_{   z}(\mathfrak{h}_4)$,  it is rather natural to  wonder whether the deformed LH systems $\>X_z$~(\ref{im}) and 
 $\widetilde{\>X}_z^3$ (\ref{jm3}) could be transformed, respectively,  into the classical one  $\>X$ (\ref{ed}) and three copies of it by means of some change of variables. In what follows we solve this question although we advance that  the answer is negative as it could be expected by taking into account  the commutation relations (\ref{in}) corresponding to a  Stefan--Sussman distribution.

Let $A$ be the trivial  Hopf algebra of a given Lie algebra $\mathfrak{g}={\rm span}\{ v_1,\dots ,v_\ell \}$  defined by the Lie brackets of $\mathfrak{g}$ and primitive coproduct map $\Delta_0(v_i)=v_i\otimes 1+1\otimes v_i  $. If $\mathfrak{g}$ admits  a quantum algebra deformation coming from a triangular (nonstandard) classical $r$-matrix $r\in \mathfrak{g}\otimes \mathfrak{g}$, then there    exists the so-called twist operator $F_z$ which is  constructed as formal power series    in  the deformation   parameter $z$
and coefficients in   $A\otimes A$ and must fulfil certain conditions~\cite{CP,Drinfelda,Drinfeld, Reshetikhin}. Explicitly, the deformed Hopf algebra $A_z$  from $A$ is given in terms of the same non-deformed  commutation relations of $A$   and a non-cocommutative  coproduct map $\Delta_z$ which is obtained from  the cocommutative one  $\Delta_0$ through 
\be
\Delta_z(v_i) =F_z \cdot \Delta_0(v_i)\cdot F_z^{-1} ,
\label{x1}
\ee
which ensures the coassociativity condition (\ref{ba}) for $\Delta_z$.   The twist operator for the  quantum algebra $U_{   z}(\mathfrak{h}_4)$ with classical $r$-matrix (\ref{ic}) is well-known and  it was formerly obtained in~\cite{Ogievetsky} (see also~\cite{Kulish} and references therein), namely
 \be
 F_z=\exp \bigl(-  v_3\otimes\log ( 1+ z v_2) \bigr) .
 \label{x2}
 \ee
  It is worth stressing that 
 $F_z$ was constructed  in~\cite{Ogievetsky} on the Borel subalgebra of   $\mathfrak{sl}(2,\mathbb R)$, which is isomorphic to the book algebra $\mathfrak{b}_2$, giving rise to the nonstandard or Jordanian quantum  $\mathfrak{sl}(2,\mathbb R)$ algebra used  
 in~\cite{BCFHL,BCFHLb} to deduce deformed $\mathfrak{sl}(2,\mathbb R)$-LH systems.
 
 Now we consider $\mathfrak{h}_4$ with Lie brackets given by (\ref{ib}) and primitive coproduct $\Delta_0$. We denote the   generators by $\tilde v_i$ to distinguish them from (\ref{if}) and (\ref{ig}). By applying (\ref{x1}) with the operator  (\ref{x2}) we obtain the quantum algebra $U_{   z}(\mathfrak{h}_4)$  with non-deformed commutation relations formally identical to (\ref{ib})  and deformed coproduct given by~\cite{twists}
\bea
&&\!\!\!\!\!\!\!\!\!\!\!\! 
  \Delta_z(\vv_2)=\vv_2\otimes 1 + 1 \otimes \vv_2+ z \vv_2\otimes \vv_2,\qquad  \Delta_z(\vv_0)=\vv_0\otimes 1 + 1 \otimes \vv_0,\nonumber\\[2pt]
&&\!\!\!\!\!\!\!\!\!\!\!\! 
\Delta_z(\vv_1)=\vv_1\otimes \frac{1}{1+z \vv_2}  +1\otimes \vv_1 + z  \vv_3 \otimes  \frac{\vv_0 }{1+z \vv_2} ,\qquad\Delta_z(\vv_3)=\vv_3\otimes \frac{1}{1+z \vv_2}  +1\otimes \vv_3 .
\label{x3}
\eea
Moreover, the invertible nonlinear map that connects both basis for $U_{   z}(\mathfrak{h}_4)$  is given by~\cite{twists}
\be
\vv_2=\frac{ \eee^{z v_2}-1}z ,\qquad v_2 = \frac 1 z \log (1+ z \vv_2)   ,\qquad \vv_l =v_l,\qquad l=0,1,3,
\label{x4}
\ee
which transforms the relations (\ref{if}) and the commutator analogues of (\ref{ig}) for $v_i$ into the coproduct (\ref{x3}) and undeformed commutators (\ref{ib}) for $\tilde v_i$ (where (\ref{ji}) has to be used).

Consequently, if we   construct   deformed $\mathfrak{h}_4$-LH systems from $U_{   z}(\mathfrak{h}_4)$ where the latter is given in the basis $\tilde v_i$ with undeformed commutators and deformed coproduct (\ref{x3}),   it is obvious that the same non-deformed  Hamiltonian vector fields $h_i$ (\ref{ee})  hold,  so that we do not actually obtain a  deformed LH system but the undeformed one $\>X$ given by (\ref{ed}). Furthermore, since $\Delta_z$ (\ref{x3}) is 
a homomorphism of the undeformed commutators (\ref{ib})  and fulfils the  coassociativity condition (\ref{ba})    by construction, the deformed prolongation $\widetilde{\> X}^{m+1}_z$ must correspond to the diagonal prolongation $\widetilde{\> X}^{m+1}$ of 
$\>X$ (\ref{ed}) to $(\mathbb{R}^2)^{m+1}$, being both equivalent by means of some change of variables. By mimicking the methodology of the previous subsection, we can explicitly compute the  Hamiltonian  functions  $h_{z,i}^{(2)}$  (\ref{de2}) but now  with the  undeformed expressions (\ref{ee}) and deformed coproduct (\ref{x3}), obtaining that
 \bea
&&  h_{z,2}^{(2)}=- x_1- x_2 + z x_1 x_2,\qquad  h_{z,0}^{(2)} =2,   
 \nonumber \\[2pt]
&&  h_{z,1}^{(2)} = \frac{ y_1(1+z x_1)}{1-z x_2}+   y_2,\qquad   h_{z,3}^{(2)} = \frac{x_1y_1}{1-z x_2}+ x_2 y_2,
\label{x5}
\eea
which has to be compared with~(\ref{jc}). These   functions give rise to a `deformed'  Hamiltonian $h_{z}^{(2)}$, despite they close the undeformed Poisson brackets (\ref{ef}) with respect to the  symplectic form $\omega=   \dd x_1\wedge \dd y_1+\dd x_2\wedge \dd y_2$.
This apparent contradiction is solved by introducing new variables $(x'_1,y'_1,x'_2,y'_2)$ such that
\bea
&& x_1=\frac{x'_1}{1-z x'_2},\qquad y_1= y'_1(1-z x'_2), \nonumber \\
&& x_2=x'_2,\qquad y_2= \frac{y'_2(1-z x'_2)- z x'_1 y'_1}{1-z x'_2}, 
\label{x6}
\eea
which defines a canonical transformation preserving $\omega$ that transforms the  Hamiltonian  functions (\ref{x5}) into
   the  non-deformed ones  $h_{i}^{(2)}$. In this way we prove that the deformation parameter $z$ is inessential since the system with $z\neq 0$ is canonically equivalent to the system with $z=0$. In other words, if we take as starting point the quantum algebra   $U_{   z}(\mathfrak{h}_4)$  in the basis with undeformed commutation rules and deforme coproduct (\ref{x3}), no distribution (\ref{bj}) arises and we recover the undeformed LH results under a suitable canonical transformation.
   
However, it can be shown by direct computation that the deformed LH systems $\>X_z$~(\ref{im}) that we have presented cannot be transformed into the undeformed system $\>X$ (\ref{ed}) through a canonical transformation. Therefore, the deformation obtained through $U_{   z}(\mathfrak{h}_4)$  in the basis $ v_i$ with deformed commutation rules turns out to be an essential one. 

In particular, if we consider the deformed  Hamiltonian functions $h_{z,i}$ (\ref{ii}) in the variables $(x,y)$, the nonlinear twist map (\ref{x4}) that relates the two basis of the quantum algebra $U_{   z}(\mathfrak{h}_4)$ induces a canonical transformation given by
\be
\tilde x = \frac{1-{\rm e}^{-z x} }{z},\qquad \tilde y = {\rm e}^{z x} y,
\label{x7}
\ee
where $\omega  =   \dd x \wedge \dd y =   \dd \tilde x \wedge \dd \tilde y$. In the new variables the  functions $h_{z,i}$ (\ref{ii}) and the system of  differential equations (\ref{im})  become
\be 
h_{z,1}=\tilde y , \qquad h_{z,2}= -x =\frac 1 z \log (1- z \tilde x),\qquad  h_{z,3}=\tilde x  \tilde y,\qquad   h_{z,0}=1,
\label{x8}
\ee
\be
 \frac{\dd \tilde x}{\dd t}= b_1(t) + b_3(t)  \tilde x,\qquad  \frac{\dd \tilde y}{\dd t}= b_2(t)\, \frac{1}{1-z\tilde x} - b_3(t) \tilde y .
\label{x9}
\ee
Thus, the nonlinear twist map (\ref{x4}) can be thought to be useful in the sense that
it provides a `minimal' LH deformation~\eqref{x9} with respect to the undeformed system (\ref{ed}), but in any case the deformation turns out to be a genuine non-trivial one, and the  functions $h_{z,i}$ (\ref{ii}) are `truly' deformed Hamiltonians.


\subsection{Deformed book Lie--Hamilton systems and Bernoulli  equations}
\label{s43}

One of the remarkable algebraic properties of the nonstandard quantum deformation of the oscillator algebra $\mathfrak{h}_4$ is that the book  subalgebra  $\mathfrak{b}_2$ remains as a Hopf subalgebra after the deformation, as can be inferred from the classical $r$-matrix in (\ref{ic}). Therefore, by construction, we obtain a Poisson sub-coalgebra spanned by $v_2$ and $v_3$ within the relations (\ref{if}) and (\ref{ig}). As a byproduct, from (\ref{ii})--(\ref{inb}) we directly get all the ingredients that characterize the resulting deformed $\mathfrak{b}_2$-LH systems, which reduce to the expressions (\ref{ga})--(\ref{gc}) under the limit $z\to 0$; these are
\be
h_{z,2}= -x,\qquad  h_{z,3}=\left( \frac{ \eee^{z x} -1}{z}\right) y,\qquad  \{h_{z,2},h_{z,3}\}_\omega=\frac{1-\eee^{-z h_{z,2}}}{z}\, ,
\label{ka}
\ee
\be
\>X_{z,2}= \frac{\partial}{\partial y} ,\qquad  \>X_{z,3}=\left( \frac{ \eee^{z x}-1}{z} \right)\frac{\partial}{\partial x}- \eee^{z x} y \, \frac{\partial}{\partial y} \, ,
\qquad  [{\bf X}_{z,2},{\bf X}_{z,3}]= - \eee^{z x} \, {\bf X}_{z,2},
\label{kb}
\ee
\bea
&&\frac{\dd x}{\dd t}=  b_3(t) \left( \frac{ \eee^{z x}-1}{z} \right) ,\nonumber\\[2pt]
&&\frac{\dd y}{\dd t}= b_2(t) - b_3(t) \eee^{z x}   y   .
\label{kc}
\eea
Consequently, the prolonged deformation of    $\mathfrak{b}_2$-LH systems   to $(\mathbb{R}^2)^3$ is straightforwardly achieved  by setting $b_1(t)\equiv 0$ in $\widetilde{\>X}_z^3$ (\ref{jm3}), or by only considering the deformed vector fields $ \> X_{h^{(3)}_{z,2}}$ and    $ \> X_{h^{(3)}_{z,3}}$  in the expressions (\ref{jn3}) and    (\ref{inx}). Moreover, the 
corresponding deformed constants of the motion and superposition rules are exactly those given  by (\ref{jn}) and (\ref{jp})   for the prolonged deformation  of $\mathfrak{h}_4$-LH systems. These results can further be  applied to   all the $\mathfrak{b}_2$-LH systems mentioned in Subsection~\ref{s32}.

To illustrate the latter point, we briefly present the main results concerning the PH deformation of the complex Bernoulli differential equations  studied in Subsection~\ref{s33}. We keep the symplectic form (\ref{hf}), the change of variables (\ref{hj}) and the relationships (\ref{hk}). It is easily seen that, in these conditions, the deformed Hamiltonian functions are given by 
\bea
&& \bar h_{z,1}= -\frac{\cos[\theta(s-1)]} {z\, r^{s-1}}\left(\exp\left\{\frac{z\, r^{s-1}}{\sin[\theta(s-1)]}     \right\}-1 \right)   ,\nonumber \\
&&
\bar h_{z,2}= -\frac{r^{s-1}}{\sin[\theta(s-1)]} ,  \qquad \{ \bar h_{z,1},\bar h_{z,2}\}_\omega=(s-1)\,\frac{ \eee^{-z \bar h_{z,2}}-1}{z}\, ,
\label{kd}
 \eea
 while the corresponding deformed vector fields $\>Y_{z,i}$ turn out to be
\bea
&& 
 {\>Y}_{z,1}=\left( r\cos^2[\theta(s-1)]  \exp\left\{  \frac{z\, r^{s-1}}{\sin[\theta(s-1)]}  \right\} + \frac{  \sin^3[\theta(s-1)]}  {z\, r^{s-2}}  \left( \exp\left\{  \frac{z\, r^{s-1}}{\sin[\theta(s-1)]}  \right\} -1  \right)\right) \frac{\partial}{\partial r}   \nonumber \\[2pt]
&& \qquad   \qquad + \sin^2[\theta(s-1)]\left( \frac{ \exp\left\{\frac{z\, r^{s-1}}{\sin[\theta(s-1)]}     \right\}}{\tan[\theta(s-1)]}- \frac{\cos[\theta(s-1)] }{z\, r^{s-1}}  \left( \exp\left\{  \frac{z\, r^{s-1}}{\sin[\theta(s-1)]}  \right\} -1  \right) \right) \frac{\partial}{\partial \theta} ,\nonumber \\[2pt]
&& 
 {\>Y}_{z,2}=r^s\cos[\theta(s-1)]\frac{\partial}{\partial r}+r^{s-1}\sin[\theta(s-1)]\frac{\partial}{\partial \theta} ,\nonumber \\[2pt]
&& 
[ {\>Y}_{z,1}, {\>Y}_{z,2}]=(s-1)   \exp\left\{  \frac{z\, r^{s-1}}{\sin[\theta(s-1)]}  \right\}  {\>Y}_{z,2} \, .
 \label{ke}
 \eea
 The deformed Bernoulli system of differential equations adopts the form
      \bea
&&\frac{\dd r}{\dd t}=  a_1(t) \left( r\cos^2[\theta(s-1)]  \exp\left\{  \frac{z\, r^{s-1}}{\sin[\theta(s-1)]}  \right\} + \frac{  \sin^3[\theta(s-1)]}  {z\, r^{s-2}}  \left( \exp\left\{  \frac{z\, r^{s-1}}{\sin[\theta(s-1)]}  \right\} -1  \right)\right) \nonumber \\[2pt]
&& \qquad\qquad   +a_2(t) r^s\cos[\theta(s-1) ],\nonumber\\[2pt]
&&\frac{\dd \theta}{\dd t}=  a_1(t)  \sin^2[\theta(s-1)]\left( \frac{ \exp\left\{\frac{z\, r^{s-1}}{\sin[\theta(s-1)]}     \right\}}{\tan[\theta(s-1)]}- \frac{\cos[\theta(s-1)] }{z\, r^{s-1}}  \left( \exp\left\{  \frac{z\, r^{s-1}}{\sin[\theta(s-1)]}  \right\} -1  \right) \right)  \nonumber\\[2pt]
&& \qquad\qquad +a_2(t)  r^{s-1} \sin[\theta(s-1) ],
\label{kf}
\eea
where the undeformed limit $z\to 0$ is given by (\ref{hb}). We also recall that it is possible to take a power series expansion in the deformation parameter $z$ in order to interpret this result as a perturbation  of the initial
 Bernoulli differential equations.

In spite of the apparently very cumbersome form of the resulting deformed Bernoulli system (\ref{kf}), its prolonged deformation along with     the  corresponding deformed  constants of the motion and    superposition rules  can explicitly be   derived from the results obtained in the   Subsection~\ref{s41}.  For the sake of brevity, we merely indicate that $F_z^{(2)}$ in (\ref{je}) now becomes 
 \be
 \begin{split}
F_z^{(2)}= & \frac 1 {z(1-s)} \left(  2-  \exp\left\{  -\frac{z\, r_1^{s-1}}{\sin[\theta_1(s-1)]}  \right\}  -  \exp\left\{  \frac{z\, r_2^{s-1}}{\sin[\theta_2(s-1)]}  \right\}  \right) \nonumber\\
& \times \left(  \frac{\cos[\theta_1(s-1)]}{r_1^{s-1}}-\frac{\cos[\theta_2(s-1)]}{ r_2^{s-1}}    \right), 
\label{kg}
\end{split}
\ee
and its undeformed limit is given in  (\ref{hl}).


\section{Concluding remarks}
\label{CR}

  In this work, a relevant question addressed to but left open in~\cite{BCFHL}, concerning the possibility of deducing a computationally feasible  deformed analogue of   superposition principles for PH deformations of LH systems, has been answered in the affirmative. This has been achieved by combining the formalism of PH deformations with the 
superintegrability property of systems having     coalgebra symmetry. In this way, two separate sets of constants of the motion have been derived for the prolonged deformations, from which a sufficient number of functionally independent constants of the motion   can be extracted, hence making it possible to establish a generic deformed superposition rule of the lowest possible order, regardless on the particular structure of the Hopf algebra deformation. This approach   amends and generalizes the construction previously proposed  in~\cite{B2013} based on permutations in the tensor product space, which  did not take into account  the symmetry breaking originated by deformed coproducts, which  prevents that a constant of the motion retains its invariant character after having been transformed by a permutation of the variables. 
It is worth stressing that in  order to develop this refinement,  it has  been necessary to introduce two new notions:   prolonged PH deformations of   LH systems and   deformed superposition rules, which respectively reduce to the usual  diagonal   prolongations and superposition rules of the   initial LH system under the non-deformed limit of the deformation parameter. Consequently, 
a complete correspondence between the characteristic properties of LH systems and their PH deformations has been established.

Along these lines, the LH systems based on the oscillator algebra $\mathfrak{h}_4$ (see \cite{Ba2015}) and their nonstandard   deformation have been studied, and an explicit deformed superposition rule for  their prolonged   deformation  has been obtained. Besides the undeniable physical interest of the oscillator algebra $\mathfrak{h}_4$, another remarkable feature has led to this choice for illustrating the generalization of the  formalism. The fact that the   book algebra $\mathfrak{b}_2$ is preserved as a Hopf subalgebra after deformation, implies that prolonged deformations of LH systems based on  $\mathfrak{b}_2$ can easily be obtained through  restriction of the prolonged deformations of $\mathfrak{h}_4$-LH systems. A striking particular case is given by the 
  prolonged deformation of complex Bernoulli equations, for which the method provides a systematic prescription for determining the constants of the motion and a deformed superposition rule.  In this context, it is worth to be mentioned that PH deformations of  LH systems based on  $\mathfrak{b}_2$,  but seen as a PH subalgebra of the nonstandard   deformation of $\mathfrak{sl}(2,\mathbb{R})$ considered in~\cite{BCFHL,BCFHLb},  have recently been used for the description of new SIS epidemic models in~\cite{CSH}. This suggests in a natural way to analyze analogous models based on $\mathfrak{b}_2$-LH systems   obtained as restrictions of prolonged PH deformations  of oscillator LH  systems. Even if under the limit $z\to 0$ the algebra $\mathfrak{b}_2$ is obviously the same, it is expected that the properties of such deformed models  should be quite distinct to those studied in \cite{CSH}, due to the different features of the $\mathfrak{sl}(2,\mathbb{R})$ and $\mathfrak{h}_4$   deformations.

The extended formalism here presented gives rise to a number of interesting questions that can be considered. A first one concerns a systematic analysis of prolonged PH deformations of LH systems in the plane, and its eventual identification with dynamical systems appearing in various applications. This in particular applies to those systems that can be interpreted as small perturbations of LH systems, and where the deformation formalism may provide a precise insight on the exact role of the deformation parameter with respect to stability or bifurcation properties of the system, as well as concerning the geometrical and dynamical behaviour of the orbits. At a more profound level, it may be asked if   PH deformations admit some kind of inverse problem. More specifically, it is conceivable that a non-autonomous parametrized nonlinear system of differential equations, without having the structure of a LH system, still allows a description in terms of a $t$-dependent vector field, the $t$-independent components of which, although not generating a finite-dimensional Lie algebra, span a distribution in the Stefan--Sussman sense. In these conditions, it would be of interest to know whether a compatible PH structure on an appropriate manifold can be found, so that one or more of the parameters in the system can be identified with   deformation parameters, hence allowing the system to be associated with a PH deformation of some LH system that would be recovered by a limiting process. This problem is intimately related to the development of an unambiguous notion of equivalence classes for PH deformations, possibly focusing on certain structural properties that until now have not been inspected in full detail, so that some kind of classification parallel to that of LH systems may be established.  Progress in some of the above-mentioned problems will hopefully be reported in some future work. 


\section*{Acknowledgements}

 \small
 
A.B.~and F.J.H.~have been partially supported by Ministerio de Ciencia e Innovaci\'on  (Spain) under grant MTM2016-79639-P (AEI/FEDER, UE), by Agencia Estatal de Investigaci\'on (Spain) under grant  PID2019-106802GB-I00/AEI/10.13039/501100011033, and by Junta de Castilla y Le\'on (Spain) under grants BU229P18 and BU091G19.   The research of R.C.S.~was financially   supported by grants MTM2016-79422-P  (AEI/FEDER, EU) and PID2019-106802GB-I00/AEI/10.13039/501100011033.
E.F.S.~acknowledges a fellowship (grant CT45/15-CT46/15) supported by the  Universidad Complutense de Madrid.\break   J. de L.~acknowledges funding from the Polish National Science Centre under grant HARMONIA 2016/22/ M/ST1/00542.


\small



\begin{thebibliography}{99}


\phantomsection
\addcontentsline{toc}{section}{References}
 
\bibitem{BCFHL} A.~Ballesteros, R.~Campoamor-Stursberg, E.~Fern\'andez-Saiz, F.J.~Herranz and J.~de Lucas.
Poisson--Hopf algebra deformations of Lie--Hamilton systems.
 {\em J. Phys. A: Math. Theor.} {\bf 51} (2018) 065202.
  \newblock \href {https://doi.org/10.1088/1751-8121/aaa090}
  {\path{doi:10.1088/1751-8121/aaa090}}
  
\bibitem{BB2015}
A.~Ballesteros, A.~Blasco, F.J.~Herranz, J.~de Lucas  and C.~Sard\'on.   Lie--Hamilton systems on the plane: properties, classification and applications. {\em J. Differ. Equ.}  {\bf 258} (2015)  2873--2907.
 \newblock \href {https://doi.org/10.1016/j.jde.2014.12.031}
  {\path{doi:10.1016/j.jde.2014.12.031}}
    
  \bibitem{BCFHLb} A.~Ballesteros, R.~Campoamor-Stursberg, E.~Fern\'andez-Saiz, F.J.~Herranz and J.~de Lucas.
A unified approach to Poisson--Hopf deformations of Lie--Hamilton systems based on $\mathfrak{sl}(2)$.
In {\em Quantum Theory and Symmetries with Lie Theory and Its Applications in Physics}, Volume 1, V. Dobrev (ed.), 
Springer Proceedings in Mathematics \& Statistics {\bf 263}, pp.~347--366.
  \newblock \href {https://doi.org/10.1007/978-981-13-2715-5_23}
  {\path{doi:10.1007/978-981-13-2715-5_23}}

\bibitem{BBHMO2009}
A.~Ballesteros, A.~Blasco, F.J.~Herranz, F.~Musso and O.~Ragnisco.  (Super)integrability from coalgebra symmetry: formalism and applications.  {\em J. Phys.: Conf. Ser.}  {\bf 175} (2009)  012004.
 \newblock \href {https://doi.org/10.1088/1742-6596/175/1/012004}
  {\path{doi:10.1088/1742-6596/175/1/012004}} 
 
  \bibitem{LSbook}
J.~de Lucas  and C.~Sard\'on.
{\em A Guide to Lie Systems with Compatible Geometric Structures.} (Singapore: World Scientific) 2020.
 \newblock \href {https://doi.org/10.1142/q0208}
  {\path{doi:10.1142/q0208}}


 \bibitem{B2013}
A.~Ballesteros, J.F.~Cari\~nena, F.J.~Herranz, J.~de Lucas  and C.~Sard\'on.   From constants of motion
to superposition rules for Lie--Hamilton systems. {\em J. Phys. A: Math. Theor.} {\bf 46} (2013) 285203.
 \newblock \href {https://doi.org/10.1088/1751-8113/46/28/285203}
  {\path{doi:10.1088/1751-8113/46/28/285203}}



\bibitem{Ba2015}
A.~Blasco, F.J.~Herranz, J.~de Lucas  and C.~Sard\'on.  Lie--Hamilton systems on the plane: applications and superposition rules. {\em  J. Phys. A: Math. Theor.} {\bf 48} (2015) 345202.
\newblock \href {https://doi.org/10.1088/1751-8113/48/34/345202}
  {\path{doi:10.1088/1751-8113/48/34/345202}}
 
 \bibitem{h4}  A.~Ballesteros and F.J.~Herranz.
Lie bialgebra quantizations of the oscillator algebra and their universal $R$-matrices.
 {\em J. Phys. A: Math. Gen.} {\bf 29} (1996) 4307-4320.
  \newblock \href {https://doi.org/10.1088/0305-4470/29/15/006}
  {\path{doi:10.1088/0305-4470/29/15/006}}

 
\bibitem{LS}
S.~Lie   and G.~Scheffers. 
{\em Vorlesungen \"uber continuierliche Gruppen mit geometrischen und anderen Anwendungen} (Leipzig: Teubner) 1893.
 
\bibitem{VES}  E.~Vessiot.   Sur quelques \'equations diff\'erentielles
ordinaires du second ordre.  {\em Annales Fac. Sci. Toulouse 1\`ere
S\'er.} {\bf 9} (1895) 1--26.
  \newblock \href {https://doi.org/10.5802/afst.117}
  {\path{doi:10.5802/afst.117}}
  
\bibitem{DAV} H.T.~Davis. {\em Introduction to Nonlinear Differential
and Integral Equations} (New York: Dover) 1962.
 
\bibitem{PW}
P.~Winternitz. 
{Lie groups and solutions of nonlinear differential equations}. In 
{\em Nonlinear Phenomena (Lectures Notes in Physics} {\bf{189}}),  K.B.~Wolf (ed.),
(New York: Springer) 1983, pp.~263--331.
 
 \bibitem{CGM00} 
J.F.~Cari{\~n}ena, J.~Grabowski and G.~Marmo. 
{\em {L}ie--{S}cheffers Systems: a Geometric Approach} (Naples: Bibliopolis) 2000.

\bibitem{CGM07} 
J.F.~Cari{\~n}ena, J.~Grabowski and G.~Marmo.  Superposition rules, Lie theorem and partial differential equations. {\em  Rep. Math. Phys.} {\bf 60}
(2007) 237--258.
 \newblock \href {https://doi.org/10.1016/S0034-4877(07)80137-6}
  {\path{doi:10.1016/S0034-4877(07)80137-6}}

 
 \bibitem{CGL2010}
 J.F.~Carine\~na,  J.~Grabowski  and J.~de Lucas.   Lie families: theory and applications. {\em J. Phys. A:
Math. Theor.} {\bf  43} (2010)  305201.
  \newblock \href {https://doi.org/10.1088/1751-8113/43/30/305201}
  {\path{doi:10.1088/1751-8113/43/30/305201}}

  \bibitem{CL2011}
 J.F.~Carine\~na  and J.~de Lucas.   Lie systems: theory, generalisations, and applications. {\em Dissertations
Math.} {\bf  479}  (2011) 1--162 (Rozprawy Mat.).
   \newblock \href {https://doi.org/10.4064/dm479-0-1}
  {\path{doi:10.4064/dm479-0-1}}


\bibitem{CGL12} 
J.F.~Cari{\~n}ena, J.~Grabowski and J.~de Lucas.  Superposition rules for higher order systems  and
their applications. {\em J. Phys. A: Math. Theor.} {\bf 45}  (2012) 185202.
 \newblock \href {https://doi.org/10.1088/1751-8113/45/18/185202}
  {\path{doi:10.1088/1751-8113/45/18/185202}}

 
\bibitem{Insel1} 
A. Inselberg. {\em On classication and superposition principles for nonlinear operators}, Thesis (Ph.D.), 
University of Illinois at Urbana-Champaign, ProQuest LLC, Ann Arbor, MI, 1965.

\bibitem{Insel2} 
A. Inselberg.  Superpositions for nonlinear operators. I. Strong superpositions and linearizability. {\em J. Math. Anal. Appl.} {\bf 40} (1972), 494--508.
 \newblock \href {https://doi.org/10.1016/0022-247X(72)90065-0}
  {\path{doi:10.1016/0022-247X(72)90065-0}}

 

\bibitem{Levin}
S.A. Levin.  Principles of nonlinear superposition. {\em J. Math. Anal. Appl.} {\bf 30} (1970) 197--205. 
 \newblock \href {https://doi.org/10.1016/0022-247X(70)90192-7}
  {\path{doi:10.1016/0022-247X(70)90192-7}}

 
\bibitem{Kon} 
B.G. Konopelchenko. Elementary B\"acklund transformations, nonlinear superposition principle and solutions of integrable equations.
{\em Phys. Lett. A}  {\bf 87} (1982), 445--448.
 \newblock \href {https://doi.org/10.1016/0375-9601(82)90754-X}
  {\path{doi:10.1016/0375-9601(82)90754-X}}

 
\bibitem{Shn} 
S. Shnider and P. Winternitz. Classification of systems of nonlinear ordinary differential equations with superposition principles.
{\em J. Math. Phys.}  {\bf 25} (1984), 3155--3165.
\newblock \href {https://doi.org/10.1063/1.526085}
  {\path{doi:10.1063/1.526085}}

 

\bibitem{Joa}
J.M. Goard and P. Broadbridge. Nonlinear superposition principles obtained by Lie symmetry methods. {\em J. Math. Anal. Appl.} {\bf  214} (1997),  633--657.
\newblock \href {https://doi.org/10.1006/jmaa.1997.5604}
  {\path{doi:10.1006/jmaa.1997.5604}}

 
\bibitem{Doro}
V.A Dorodnitsyn. The non-autonomous dynamical systems and exact solutions with superposition principle for evolutionary PDEs. {\em Ufimsk.~Mat.~Zh.} {\bf 4} (2012), 186--195.
\newblock \href {http://mi.mathnet.ru/eng/ufa180}
  {\path{http://mi.mathnet.ru/eng/ufa180}}
 


 \bibitem{CS2016}
R.~Campoamor-Stursberg.  Low dimensional Vessiot--Guldberg Lie algebras of second-order ordinary differential equations. {\em Symmetry} {\bf  8} (2016) 8030015.
 \newblock \href {https://doi.org/10.3390/sym8030015}
  {\path{doi:10.3390/sym8030015}}


 \bibitem{CS2016b}
   R.~Campoamor-Stursberg.  A functional realization of $\mathfrak{sl}(3, \mathbb R)$ providing minimal Vessiot--Guldberg--Lie algebras of nonlinear second-order ordinary differential equations as proper subalgebras. {\em  J. Math. Phys.} {\bf  57} (2016) 063508.
    \newblock \href {https://doi.org/10.1063/1.4954255}
  {\path{doi:10.1063/1.4954255}}

 
    \bibitem{Ibragimov2016}
  N.H.~Ibragimov  and A.A.~Gainetdinova.  Three-dimensional dynamical systems admitting nonlinear superposition with three-dimensional Vessiot--Guldberg--Lie algebras. {\em Appl. Math. Lett.} {\bf  52} (2016) 126--131.
 \newblock \href {https://doi.org/10.1016/j.aml.2015.08.012}
  {\path{doi:10.1016/j.aml.2015.08.012}}


 


\bibitem{CLS2013} 
J.F.~Cari{\~n}ena,  J.~de Lucas and C.~Sard\'on.  Lie--Hamilton systems: theory and applications. {\em Int. J.
Geom. Methods Mod. Phys.} {\bf 10} (2013) 1350047.
 \newblock \href {https://doi.org/10.1142/S0219887813500473}
  {\path{doi:10.1142/S0219887813500473}}




 \bibitem{HLT}  F.J.~Herranz, J.~de Lucas and M.~Tobolski.
Lie--Hamilton systems on curved spaces: a geometrical approach.
 {\em J. Phys. A: Math. Theor.} {\bf 50} (2017) 495201.
  \newblock \href {https://doi.org/10.1088/1751-8121/aa918f}
  {\path{doi:10.1088/1751-8121/aa918f}}



 \bibitem{Vaisman}  I.~Vaisman.  {\em Lectures on the Geometry of Poisson manifolds} (Progress in Mathematics vol.~118)
(Basel: Birkh\"auser Verlag) 1994.


 \bibitem{CP}   V.~Chari and A.~Pressley. {\em A Guide to Quantum Groups}. (Cambridge: Cambridge University Press) 1994.
 
 
\bibitem{majid}   S.~Majid. {\em Foundations of Quantum Group Theory}.  (Cambridge: Cambridge University Press)    1995.

 \bibitem{Abe}   E.~Abe. {\em Hopf Algebras},  Part of Cambridge Tracts in Mathematics. (Cambridge: Cambridge University Press)    2004.
 

\bibitem{Pa57}
 R.S.~Palais.  
{\em A global formulation of the Lie theory of transformation groups}. 
{Memoirs American Math. Soc. {\bf 22}}  (Providence RI: American Mathematical Society) 1957.

  \bibitem{WA} J.F.~Cari{\~n}ena, A.~Ibort, G.~Marmo  and G.~Morandi.  {\em Geometry from Dynamics, Classical and Quantum} (New York: Springer) 2015.
  \newblock \href {https://doi.org/10.1007/978-94-017-9220-2}
  {\path{doi:10.1007/978-94-017-9220-2}}

\bibitem{BCR96}
A.~Ballesteros, M.~Corsetti and O.~Ragnisco.  $N$-dimensional classical integrable systems from Hopf algebras.  {\em Czech. J. Phys.}  {\bf 46} (1996)  1153--1163.
 \newblock \href {https://doi.org/10.1007/BF01690329}
  {\path{doi:10.1007/BF01690329}}

\bibitem{BR98}
A.~Ballesteros and O.~Ragnisco.  A systematic construction of completely integrable Hamiltonians from coalgebras.  {\em J. Phys. A: Math. Gen.}  {\bf 31} (1998)  3791--3813.
 \newblock \href {https://doi.org/10.1088/0305-4470/31/16/009}
  {\path{doi:10.1088/0305-4470/31/16/009}}
  
  
   \bibitem{BHMO2004}
A.~Ballesteros, F.J.~Herranz, F.~Musso and O.~Ragnisco.  Superintegrable deformations of the Smorodinsky--Winternitz Hamiltonian.  
In {\em Superintegrability in Classical and Quantum Systems}, P.~Tempesta {\em et al} (eds.), CRM Proc. and Lecture Notes, vol.~37
(Providence, RI: American Mathematical Society), 2004, pp.~1--14.
 \newblock \href {https://doi.org/10.1090/crmp/037/01}
  {\path{doi:10.1090/crmp/037/01}}



   \bibitem{BH2007}
A.~Ballesteros and F.J.~Herranz.  Universal integrals for superintegrable systems on $N$-dimensional spaces of constant curvature.  {\em J. Phys. A: Math. Theor. }  {\bf 40} (2007)  F51--F59.
 \newblock \href {https://doi.org/10.1088/1751-8113/40/2/F01}
  {\path{doi:10.1088/1751-8113/40/2/F01}}


\bibitem{GKO}
A.~Gonz\'alez-L\'opez, N.~Kamran   and P.J.~Olver. Lie algebras of vector fields in the real plane.  {\em Proc. London Math. Soc.}  {\bf 64} (1992)  339--368.
 \newblock \href {https://doi.org/10.1112/plms/s3-64.2.339}
  {\path{doi:10.1112/plms/s3-64.2.339}}
 
\bibitem{Buchdahl}
H.A.~Buchdahl.  A relativistic fluid sphere resembling the Emden polytrope of index 5.  {\em Astrophys. J.}  {\bf 140} (1964)  1512--1516.
 \newblock \href {https://doi.org/10.1086/148055}
  {\path{doi:10.1086/148055}}

\bibitem{Chandrasekar}
V.K.~Chandrasekar, M.~Senthilvelan and M.~Lakshmanan.  On the complete integrability and linearization of certain second-order nonlinear ordinary differential equations.  {\em Proc. R. Soc. A}  {\bf 461} (2005)  2451--2477.
 \newblock \href {https://doi.org/10.1098/rspa.2005.1465}
  {\path{doi:10.1098/rspa.2005.1465}}

 \bibitem{Nikiciuk}
J.L.~Cie\'sli\'nski and T.~Nikiciuk.  A direct approach to the construction of standard and non-standard Lagrangians for dissipative-like dynamical systems with variable coefficients.  {\em J. Phys. A: Math. Theor.}  {\bf 43} (2010)  175205.
 \newblock \href {https://doi.org/10.1088/1751-8113/43/17/175205}
  {\path{doi:10.1088/1751-8113/43/17/175205}}

\bibitem{Tsvetkov}
D.P.~Tsvetkov.  A periodic Lotka--Volterra system.  {\em Serdica Math. J.}  {\bf 22} (1996)  109--116.
 \newblock \href {http://www.math.bas.bg/serdica/1996/1996-109-116.pdf}
  {\path{journalpdf}}

 
 \bibitem{Jin}
Z.~Jin, H.~Maoan and L.~Guihua.  The persistence in a Lotka--Volterra competition systems with
impulsive.  {\em Chaos Solitons Fractals}  {\bf 24} (2005)  1105--1117.
 \newblock \href {https://doi.org/10.1016/j.chaos.2004.09.065}
  {\path{doi:10.1016/j.chaos.2004.09.065}}

\bibitem{Muriel}
C.~Muriel and J.L.~Romero. $\lambda$-symmetries of some chains of ordinary differential equations.  {\em Nonlinear Anal.: Real World Appl.}  {\bf 16} (2014)  191--201.
 \newblock \href {https://doi.org/10.1016/j.nonrwa.2013.09.018}
  {\path{doi:10.1016/j.nonrwa.2013.09.018}}

\bibitem{Zoladek}
H.~\.Zo\l \c adek. The method of holomorphic foliations in planar periodic systems: the case of Riccati equations.  {\em J. Differ. Equ.}  {\bf 165} (2000)  143--173.
 \newblock \href {https://doi.org/10.1006/jdeq.1999.3721}
  {\path{doi:10.1006/jdeq.1999.3721}}

 \bibitem{Marino}
A.~Marino. Topological methods, variational inequalities and elastic bounce trajectories.  {\em Rend. Lincei Mat. Appl.}  {\bf 22} (2011)  269--290.
 \newblock \href {https://doi.org/10.4171/RLM/600}
  {\path{doi:10.4171/RLM/600}}


\bibitem{boson}  A.~Ballesteros, F.J.~Herranz and J.~Negro.
Boson representations, non-standard quantum algebras and contractions.
 {\em J.~Phys.~A: Math.~Gen.}~{\bf 30} (1997) 6797--809.
  \newblock \href {https://doi.org/10.1088/0305-4470/30/19/018}
  {\path{doi:10.1088/0305-4470/30/19/018}}

\bibitem{boson2}  A.~Ballesteros, F.J.~Herranz and P.~Parashar.
A Jordanian quantum two-photon/Schr\"odinger algebra.
 {\em J. Phys. A: Math. Gen.} {\bf 30} (1997) 8587--8597.
  \newblock \href {https://doi.org/10.1088/0305-4470/30/24/019}
  {\path{doi:10.1088/0305-4470/30/24/019}}
  
   
    \bibitem{Drinfelda} V.G.~Drinfel'd.
Constant quasiclassical solutions of the Yang--Baxter quantum equation.
 {\em  Dokl.  Akad. Nauk SSSR.}  {\bf 273} (1983)  531--535.
\newblock \href {http://mi.mathnet.ru/eng/dan9865}
 {\path{http://mi.mathnet.ru/eng/dan9865}}

  

  \bibitem{Drinfeld} V.G.~Drinfel'd.
Quasi-Hopf algebras.
 {\em  Leningrad Math. J.}  {\bf 1} (1990)  1419--1457.
\newblock \href {http://mi.mathnet.ru/eng/aa53}
 {\path{http://mi.mathnet.ru/eng/aa53}}

 

\bibitem{Reshetikhin} N.~Reshetikhin.
Multiparameter quantum groups and twisted quasitriangular Hopf algebras.
 {\em  Lett. Math. Phys.}  {\bf 20} (1990)  331--335.
\newblock \href {https://doi.org/10.1007/BF00626530}
 {\path{doi.org/10.1007/BF00626530}}

   \bibitem{Ogievetsky} O.~Ogievetsky.
Hopf structures on the Borel subalgebra of $sl(2)$.
 In {\em  Proc.~Winter School ``Geometry and Physics"}, J.~Bure\v{s}   and V.~Sou\v{c}ek (eds.), {\em  Rendiconti Cir. Math. Palermo, ser.~II, suppl.}  {\bf 37} (1994) 185--199.
  \newblock \href {https://dml.cz/handle/10338.dmlcz/701555}
  {\path{https://dml.cz/handle/10338.dmlcz/701555}}

  \bibitem{Kulish}  P.P.~Kulish  and A.A.~Stolin.
Deformed yangians and integrable models.
 {\em Czech. J. Phys .} {\bf 47} (1997) 1207--1212.
  \newblock \href {https://doi.org/10.1023/A:1022869414679}
  {\path{doi:10.1023/A:1022869414679}}




\bibitem{twists}  A.~Ballesteros, F.J.~Herranz, J.~Negro and L.M.~Nieto.
Twist maps for non-standard quantum algebras and discrete Schr\"odinger symmetries.
 {\em J. Phys. A: Math. Gen.} {\bf 33} (2000) 4859--4870.
  \newblock \href {https://doi.org/10.1088/0305-4470/33/27/303}
  {\path{doi:10.1088/0305-4470/33/27/303}}

\bibitem{CSH}  O.~Esen, E.~Fern\'andez-Saiz, C.~Sard\'on and M.~Zaj\c{a}c. 
Geometry and solutions of an epidemic SIS model permitting fluctuations and quantization. 
 \newblock \href {https://arxiv.org/abs/2008.02484}
  {\path{arXiv:2008.02484}}

  
\end{thebibliography}
\end{document}